\documentclass[prd,twocolumn,nofootinbib]{revtex4-1}

\usepackage{graphicx}
\usepackage{dcolumn}
\usepackage{bm}
\usepackage[colorlinks=true]{hyperref}
\usepackage[]{amsmath}
\usepackage{amssymb}
\usepackage{pifont}
\newcommand{\xmark}{\ding{55}}
\newcommand{\Msolar}{\;\text{\(\textup{M}_\odot\)}}
\usepackage{tabls}
\usepackage{soul}
\usepackage{gensymb}
\usepackage{multirow}
\usepackage[dvipsnames]{xcolor}

\newcommand{\GR}{{\mbox{\tiny GR}}}
\newcommand{\KG}{{\mbox{\tiny kh}}}
\newcommand{\CG}{{\mbox{\tiny CG}}}
\newcommand{\EA}{{\mbox{\tiny EA}}}
\newcommand{\ST}{{\mbox{\tiny ST}}}
\newcommand{\GRk}{{\mbox{\tiny GR,k}}}
\newcommand{\dCS}{{\mbox{\tiny dCS}}}
\newcommand{\EdGB}{{\mbox{\tiny EdGB}}}
\newcommand{\Q}{{\mbox{\tiny Q}}}
\newcommand{\VEV}{{\mbox{\tiny VEV}}}
\newcommand{\SMG}{{\mbox{\tiny SMG}}}
\newcommand{\PL}{{\mbox{\tiny PL}}}
\newcommand{\BH}{{\mbox{\tiny BH}}}

\begin{document}

\title{Testing General Relativity with Black Hole-Pulsar Binaries}

\author{Brian Seymour}
\affiliation{Department of Physics, University of Virginia, Charlottesville, Virginia 22904, USA}

\author{Kent Yagi}
\affiliation{Department of Physics, University of Virginia, Charlottesville, Virginia 22904, USA}

\date{\today}

\begin{abstract} 
Binary pulsars allow us to carry out precision tests of gravity and have placed stringent bounds on a broad class of theories beyond general relativity. 
Current and future radio telescopes, such as FAST, SKA, and MeerKAT, may find a new astrophysical system, a pulsar orbiting around a black hole, which will provide us a new source for probing gravity.
In this paper, we systematically study the prospects of testing general relativity with such black hole-pulsar binaries. 
We begin by finding a mapping between generic non-Einsteinian parameters in the orbital decay rate and theoretical constants in various modified theories of gravity and then summarize this mapping with a ready-to-use list. 
Theories we study here include scalar-tensor theories, varying $G$ theories, massive gravity theories, generic screening gravity and quadratic curvature-corrected theories.
We next use simulated measurement accuracy of the orbital decay rate for black hole-pulsar binaries with FAST/SKA and derive projected upper bounds on the above generic non-Einsteinian parameters.
We find that such bounds from black hole-pulsars can be stronger than those from neutron star-pulsar and neutron star-white dwarf binaries by a few orders of magnitude when the correction enters at negative post-Newtonian orders. 
By mapping such bounds on generic parameters to those on various modified theories of gravity, we find that one can constrain the amount of time variation in Newton's constant $G$ to be comparable to or slightly weaker than than the current strongest bound from solar system experiments, though the former bounds are complementary to the latter since they probe different regime of gravity.
We also study how well one can probe quadratic gravity from black hole quadrupole moment measurements of black hole-pulsars.
We find that bounds on the parity-violating sector of quadratic gravity can be stronger than current bounds by six orders of magnitude.
These results suggest that a new discovery of black hole-pulsars in the future will provide powerful ways to probe gravity further.

\end{abstract}
\maketitle

\section{Introduction}
\label{sec:intro}


General relativity (GR) is currently the most well-tested theory of gravity. Nevertheless, due to its inconsistency with quantum mechanics, it is only an effective field theory below some energy threshold. Eventually, at some length or energy scales, one expects to find deviations from GR predictions. So far, GR has passed all the tests with flying colors~\cite{Will:2014kxa}. Thus, it is important to probe even deeper to search for non-GR effects using new sources that we hope to detect with current and future detectors. Probing GR can also shed light on cosmology as another motivation to consider going beyond GR is to explain current accelerating expansion of our universe and missing mass problem in galaxies without introducing dark energy or dark matter~\cite{Jain:2010ka,Clifton:2011jh,Joyce:2014kja,Koyama:2015vza}.

Various probes of gravity can be classified into weak-field and strong-field tests~\cite{Psaltis:2008bb,Baker:2014zba,Yunes:2016jcc}. Solar system experiments~\cite{Will:2014kxa} and cosmological observations~\cite{Jain:2010ka,Clifton:2011jh,Joyce:2014kja,Koyama:2015vza} fall into the weak-field category, as the amount of curvature and gravitational potential that these experiments and observations probe is weak and small. On the other hand, recent direct gravitational-wave (GW) measurements from binary black hole (BH) mergers~\cite{Abbott:2016blz,Abbott:2016nmj,TheLIGOScientific:2016pea,Abbott:2017vtc,Abbott:2017oio,Abbott:2017gyy} allow us to probe the strong and dynamical field regime of gravity for the first time~\cite{TheLIGOScientific:2016src,Yunes:2016jcc}.

Precision tests of gravity have also been carried out via binary pulsar (PSR)\footnote{In this paper, binary PSR refers to a binary consisting of a PSR with a \emph{stellar} companion.} observations~\cite{stairs,WexRadioPulsars} (see e.g.~\cite{Archibald:2018oxs} for a very recent work on testing the strong equivalence principle with the triple system). Such systems allow us to probe both the weak-field and strong-field effects. This is because these binaries are widely-separated relative to GW sources of compact binaries that are about to coalesce (and hence weak-field), and at the same time, PSRs are neutron stars which are very compact objects (and thus strong-field). Such systems (PSRs with white dwarf (WD) companions in particular) are ideal for probing scalar dipole emission in scalar-tensor theories~\cite{Freire:2012mg} that is absent in GR. One can use parameterized post-Keplerian (PPK) parameters, such as the advance rate of periastron, orbital decay rate and Shapiro time delay, to carry out precision tests of GR. Measurements of any two PPK parameters determine the masses of compact objects in binary PSRs, while any additional PPK parameter measurements further probe the consistency of gravitational theory.

Binary BHs and binary neutron stars have been found via electromagnetic-wave~\cite{stairs,Valtonen:2008tx,WexRadioPulsars} and GW observations~\cite{Abbott:2016blz,Abbott:2016nmj,TheLIGOScientific:2016pea,Abbott:2017vtc,Abbott:2017oio,Abbott:2017gyy,TheLIGOScientific:2017qsa}. Interestingly, a binary that consists of one BH and one neutron star has not been found yet. One possibility to look for such systems is through GW observations, which are useful for probing non-GR theories such as scalar-tensor theories~\cite{berti-buonanno,Yagi:2009zm,Yagi:2009zz,Zhang:2017sym}. 

Radio observations offer another possible way of finding a PSR orbiting around a BH.
Such a system may be found with either the Five-hundred-meter Aperture Spherical radio Telescope (FAST) that is undergoing commissioning, MeerKAT or a next-generation radio telescopes such as the Square Kilometer Array (SKA). Population synthesis suggests that there may be around 3--80 BH-PSRs in the Galactic disk and FAST may detect up to 10\% of them~\cite{Shao:2018qpt}. Radio telescopes may detect two different types of PSR in the BH-PSR binary. A normal PSR is a younger, slower spinning PSR. A millisecond PSR (MSP) is older and recycled. MSPs are ideal for testing GR because they spin significantly faster and achieve a better timing precision than normal PSRs (100ns vs 100$\mu$s respectively) \cite{QuadrupoleMeasureabilitySgrA_PsaltisWex}. This better timing precision allows for a more accurate system measurement. Note that MSPs are usually harder to find than normal pulsars due to selection effects. This is particularly true for the Galactic center, where the large distance and scattering effects in the interstellar medium may have prevented a discovery so far.
Regarding formation of a BH-MSP binary, at least two possible  scenarios exist. First, an exchange interaction of a binary can create a BH-MSP binary inside a dense region such as a globular cluster or the Galactic center. Second, a BH-MSP binary can evolve directly from a finely tuned initial system of main sequence stars. For example, when the masses of initial stars are comparable, a neutron star can form first, which is being accreted by a companion and spun up, and eventually the companion collapses to a BH. Thus, while binaries with a BH and normal PSR are more likely, BH-MSP binaries may still be found.

BH-PSR systems are powerful for testing GR, including no-hair properties of BHs~\cite{QuadrupoleBoundFunctionofPb,WexBHP,QuadrupoleMeasureabilitySgrA_PsaltisWex}, scalar-tensor theories~\cite{WexBHP,Wex_ScalarTensor}, higher-curvature theories~\cite{EDGBBoundComparison}, higher-dimensional gravity~\cite{Simonetti:2010mk} and quantum gravity effects~\cite{Estes:2016wgv}. 
The reason is as follows. The relative velocity of a binary is given by $v = (2 \pi M / P)^{1/3}$ with $c=G=1$, where $M$ is the total mass while $P$ is the orbital period. For a PSR orbiting around a stellar-mass BH, $M$ is larger than that of PSR binaries with NS or WD companions, but one expects $P$ to be also larger. 
In fact, the BH-PSR relative velocity is smaller than the NS-PSR or PSR-WD case (typically by a factor of 2) because the longer period more than compensates for the larger total mass. 
Additionally, the measurement accuracy of the orbital decay rate is expected to be similar to that of NS-PSR or PSR-WD. Thus, BH-PSR systems will have more advantage on probing non-GR effects that enter at a negative post-Newton (PN) order\footnote{PN expansion assumes that the orbital motion is sufficiently slow relative to the speed of light. A correction term is said to be of relative $n$PN order when it is proportional to $(v/c)^{2n}$ relative to the leading contribution in GR.}, such as scalar dipole radiation in scalar-tensor theories.


In this paper, we study in more detail how well one can probe 
modifications to GR in theories that have not yet been studied in the context of BH-PSRs.
The first half of the paper focuses on using the orbital decay rate measurement. We will first introduce a generic parameterization that captures the non-GR modifications to the orbital decay rate~\cite{Yunes:2010qb}. We will next derive projected bounds on this parameter from a BH-MSP with FAST and SKA based on a simulated measurement accuracy in~\cite{WexBHP}. (This simulation will be discussed further in Sec.~\ref{sec:Pdot-general}.) We will then compare such bounds to those on parameterized post-Einsteinian (PPE)~\cite{Yunes:2009ke} parameters, which capture the non-GR modifications in gravitational waveforms from compact binary mergers. The PPE formalism (or its modified version) has already been applied to recent GW events~\cite{TheLIGOScientific:2016src,KentGWbounds}. We will further map such generic bounds to those on specific modified theories of gravity by creating a ``dictionary'' between the generic and theoretical parameters shown in Table~\ref{tab:GammaValues}. In particular, we will study theories with time varying gravitational constant~\cite{Will:2014kxa}, Lorentz-violating graviton mass~\cite{Finn:2001qi}, Lorentz-preserving graviton mass~\cite{MassiveGravityCG}, and generic screening mechanisms~\cite{Zhang:2018dxi}.

\newcommand{\minitab}[2][l]{\begin{tabular}{#1}#2\end{tabular}}
\renewcommand{\arraystretch}{1.6}
\begingroup
\squeezetable
\begin{table*}[htb]
\begin{centering}
\begin{tabular}{c|c|c|c|c|c|c}
\hline 
\hline 
\multirow{2}{*}{Theory}&\multirow{2}{*}{$\gamma$}&\multirow{2}{*}{$f(e)$}&\multirow{2}{*}{$n$}&Theoretical&\multirow{2}{*}{Refs.}&Stronger\\
 & & & & parameters & & bounds? \\ 
 \hline \hline
\multirow{2}{*}{Time-Varying G (Sec.~\ref{subsec:VaryingG})}&\multirow{2}{*}{ 
 $\frac{5 \dot G M^3}{48 m_{c} m_{p}} \left[1-s_{p}^{\dot G} \left(1+\frac{m_{c}}{M}\right)-s_{c}^{\dot G} \left(1+\frac{m_{p}}{M}\right)\right]f(e)$ 
 }&\multirow{2}{*}{$\frac{1}{F_\GR(e)}$}& \multirow{2}{*}{$-4$} &\multirow{2}{*}{$\dot G / G$ }&\multirow{2}{*}{\cite{WexRadioPulsars}} &\multirow{2}{*}{\checkmark} \\ 
 & & & & && \\ \hline
Lorentz-violating & \multirow{2}{*}{ $\frac{5M^2}{24 } m_g ^2 f(e)$}&\multirow{2}{*}{ $\frac{1}{(1-e^2)^{1/2}F_\GR(e)}$}&\multirow{2}{*}{$-3$}&\multirow{2}{*}{$m_g$}& 
\multirow{2}{*}{\cite{Finn:2001qi}}& \multirow{2}{*}{\xmark} \\ 
Massive Gravity (Sec.~\ref{subsec:FierzPauli}) & & & & && \\ \hline
 Cubic Galileon & \multirow{2}{*}{$\frac{25}{32} \pi \lambda ^2 \frac{ M_\PL M_\Q^2 M^{3}}{m_p^2 m_c^2} m_g f(e)$}&\multirow{2}{*}{$\frac{F_\CG(e)}{F_{\GR}(e)}$}&\multirow{2}{*}{$-11/4$}&\multirow{2}{*}{$m_g$}&\multirow{2}{*}{\cite{MassiveGravityCG}}& \multirow{2}{*}{\xmark} \\
 Massive Gravity (Sec.~\ref{subsec:CubicGalileon}) & & & & & & \\ \hline
General Screen Modified & \multirow{2}{*}{$\frac{5}{192} (\epsilon_p-\epsilon _c)^2 f(e)$} & \multirow{2}{*}{$\frac{F_\SMG} {F_\GR(e)}$}& \multirow{2}{*}{$-1$}& \multirow{2}{*}{$\phi_\VEV / M_\PL$}& \multirow{2}{*}{\cite{Zhang:2018dxi}}& \multirow{2}{*}{\xmark} \\ 
Gravity (Sec.~\ref{subsec:SMG}) & & & & & & \\ \hline
 \multirow{2}{*}{(massless) Scalar-Tensor} & \multirow{2}{*}{$ \frac{5}{96} \left(\bar\alpha_p^\ST-\bar\alpha_c^\ST\right)^2 f(e)$} & \multirow{2}{*}{$\frac{F_\SMG}{F_\GR(e)}$}& \multirow{2}{*}{$-1$} & \multirow{2}{*}{($\alpha_0$, $\beta_0$)}& \multirow{2}{*}{\cite{Freire:2012mg}} & \multirow{1}{*}{\checkmark} \\ 
& & & & & & \cite{WexBHP,Wex_ScalarTensor} \\ \hline
\multirow{2}{*}{Einstein-dilaton Gauss-Bonnet} & \multirow{2}{*}{$\frac{5 \pi}{24} \left(\bar\alpha_p^\EdGB-\bar\alpha_c^\EdGB\right)^2f(e)$}&\multirow{2}{*}{---}&\multirow{2}{*}{$-1$} &\multirow{2}{*}{$\sqrt{\alpha_\EdGB}$}&\multirow{2}{*}{\cite{EDGBBoundComparison}} & \multirow{1}{*}{\checkmark} \\ 
 & & & & & & \cite{EDGBBoundComparison} \\ \hline
 \multirow{2}{*}{Einstein-\AE ther} & \multirow{2}{*}{$ \frac{5 \mathcal{C}_\EA }{32 } \left(1-\frac{c_{14}}{2}\right) \left(s_{p}^\EA-s_{c}^\EA \right)^2f(e)$} & \multirow{2}{*}{---} &\multirow{2}{*}{$-1$}&\multirow{2}{*}{($c_{+}$, $c_{-})$}&\multirow{2}{*}{\cite{Yagi:2013ava}} & \multirow{2}{*}{?} \\ 
& & & & & & \\ \hline
 \multirow{2}{*}{Khronometric} & \multirow{2}{*}{$ \frac{5 \mathcal{C}_\KG}{32 } \left(1-\frac{\alpha_\KG}{2}\right) \left(s_{p}^\KG-s_{c}^\KG\right)^2 f(e)$} & \multirow{2}{*}{---} & \multirow{2}{*}{$-1$}& \multirow{2}{*}{($\lambda_\KG$, $\alpha_\KG$, $\beta_\KG$)}& \multirow{2}{*}{\cite{Yagi:2013ava}}& \multirow{2}{*}{?} \\ 
& & & && & \\ 
\hline
\hline 
\end{tabular} 
\end{centering}
\caption{
 \label{tab:GammaValues} 
 Mapping between non-GR parameters ($\gamma$ and $n$) in the orbital decay rate $\dot P$ in Eq.~\eqref{eq:generic-Pdot} to theoretical parameters in various example modified theories of gravity, together with some references. These expressions are valid for any compact binaries (not specific to BH-PSRs). The first four theories are those considered in Sec.~\ref{sec:Pdot-example}, while the last four theories are presented only for reference. Note that we study bounding EdGB gravity in Sec.~\ref{subsec:EdGB} via BH quadrupole moment measurement which is different from the orbital decay rate presented here. Theoretical parameters are presented in the fifth column. The last column shows whether BH-PSR bounds are stronger than other existing bounds (\checkmark: yes; \xmark: no; ?: unknown). The meaning of each parameter in the second column is as follows. $m_p$: primary PSR's mass, $m_c$ companion's mass, $M$: total system mass, $e$: eccentricity, $M_\PL$: Planck mass, $M_\Q$: a mass parameter in Eq.~\eqref{eq:MQdef}, $\lambda$: a numerical constant in Eq.~\eqref{eq:lambdadef}, $\epsilon_A$: a screening parameter in SMG in Eq.~\eqref{eq:ScreenScalarCharge}, $C_\EA$ and $C_\KG$: a function of theory parameters in Eqs.~(114) and (124) of~\cite{Yagi:2013ava} for Einstein-\ae ther and khronometric theories respectively, $c_{14}$ and $\alpha_\KG$: a combination of coupling constants in Einstein-\ae ther theory and khronometric theory respectively. $\bar\alpha_A$ is the scalar charge. In many of scalar-tensor theories, it is non-vanishing for stars while it is zero for a BH~\cite{Hawking:1972qk,Bekenstein:1995un,Sotiriou:2011dz}. In EdGB gravity, such a charge vanishes for stars~\cite{Yagi:2011xp,EDGBBoundComparison} while that for a BH is in Eq.~(37) of~\cite{Berti:2018cxi}. $s_A$ is the sensitivity and that for a neutron star in Einstein-\ae ther and khronometric theory has been computed in~\cite{Yagi:2013qpa,Yagi:2013ava}\footnote{The fitting function for the NS sensitivity in Einstein-\ae ther and khronometric theory can be found in Eq.~(186) or~(C1) of~\cite{Yagi:2013ava}, though the parameter region in which the fit is valid has mostly been ruled out by GW170817~\cite{Gumrukcuoglu:2017ijh,Oost:2018tcv}. } while that for a BH has not been calculated yet. 
The eccentricity dependent function $f(e)$ is presented in the third column if known, while ``---'' means that the correction has been calculated only for circular binaries ($f=1$). $F_\GR$ is the eccentricity dependence in GR in Eq.~\eqref{eq:FGR}, while $F_\CG(e)$ and $F_\SMG(e)$ are that in cubic Galileon massive gravity and generic screened massive gravity defined in Eqs.~\eqref{eq:FCGdef} and~\eqref{eq:F_SMG} respectively.
} 
\end{table*}
\endgroup

The second half of this paper focuses on using the BH quadrupole moment measurement with BH-PSRs to probe gravity. A non-vanishing quadrupole moment causes a periodic variation in the PSR motion~\cite{Wex:1998wt}, which can be extracted from the Roemer time delay measurement. For stellar-mass BH-MSP systems, one may be able to measure the BH quadrupole moment within ~10\% accuracy~\cite{WexBHP}, and the accuracy may be 10\% if one finds a PSR orbiting around Sgr~A$^*$~\cite{QuadrupoleBoundFunctionofPb,QuadrupoleMeasureabilitySgrA_PsaltisWex}. (The simulations used for these are discussed in~Sec.~\ref{sec:Q-measurement}.) We apply these projected measurements to quadratic curvature theories, namely Einstein-dilaton Gauss-Bonnet (EdGB) gravity~\cite{Metsaev:1987zx,Maeda:2009uy} and dynamical Chern-Simons (dCS) gravity~\cite{Jackiw:2003pm,Alexander:2009tp} for the even-parity and odd-parity sector respectively. 
Both theories are motivated from string theory. Analytic BH solutions with arbitrary spin in these theories have not been found yet. Non-rotating and slowly-rotating analytic BH solutions have been constructed in~\cite{Mignemi:1992pm,Yunes:2011we,Sotiriou:2014pfa,EDGBBoundSpin2,EDGBBoundSpin4} for EdGB and in~\cite{Yunes:2009hc,Konno:2009kg,DCSBoundSpin2,DCSBoundspin4} for dCS.

\subsection{Executive Summary}
\begin{figure}[t]
\begin{center}
\includegraphics[width=\linewidth]{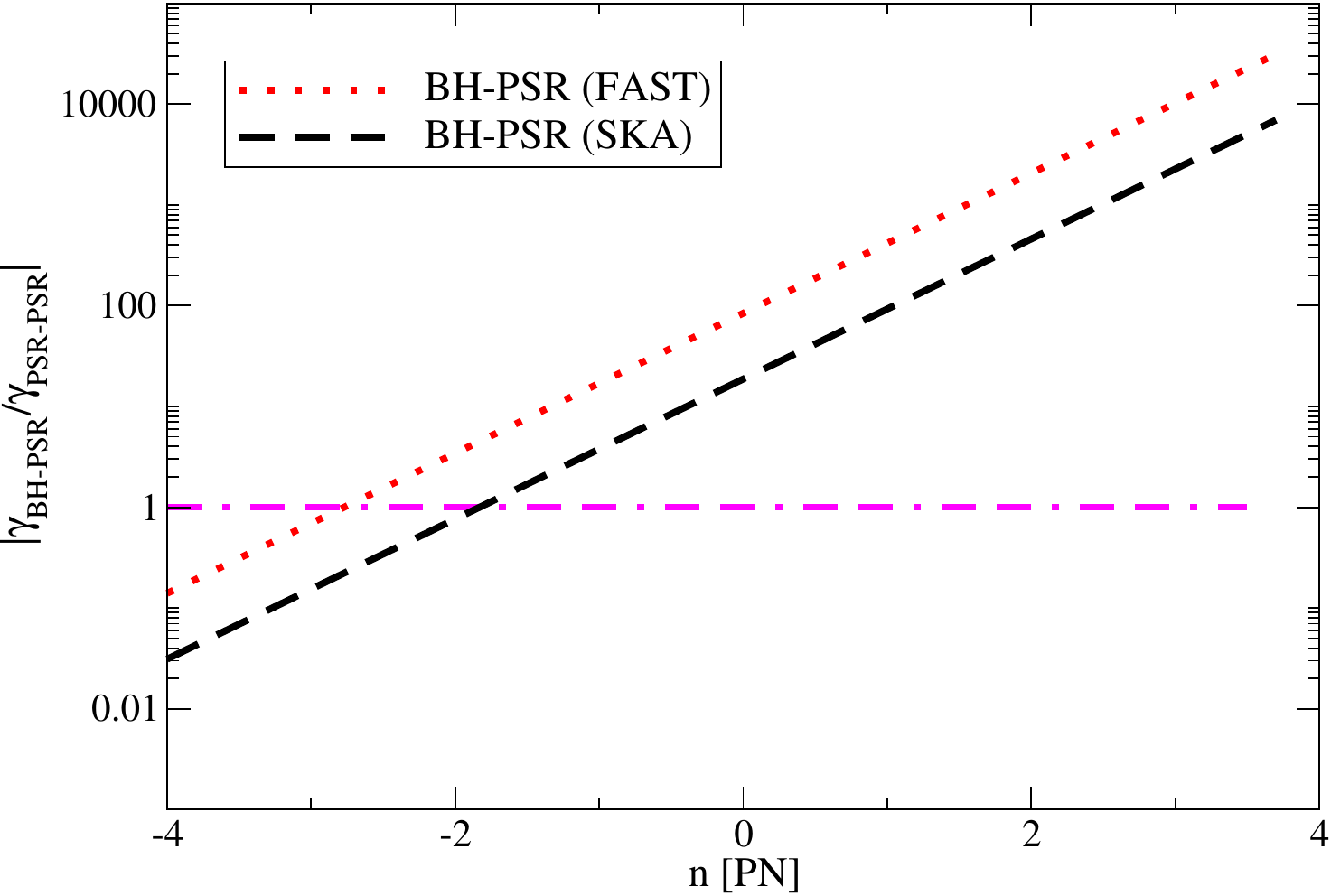}
\caption{\label{fig:gammaratio} The ratio in the upper bound on the fractional non-GR correction $\gamma$ to the orbital decay rate between BH-PSR and double PSR systems, as a function of at which PN order the correction enters. This bound uses a millisecond PSR-BH binary with simulations from \cite{WexBHP}. This figure shows how much improvement one finds by using the BH-PSR system compared to the double PSR one in terms of testing GR. For example, if the ratio is below unity (horizontal magenta dotted dashed line), the bound from the former is stronger than that from the latter. The ratio is shown for FAST (red dotted) and SKA (black dotted). Notice that FAST (SKA) has a bound that is an order of magnitude stronger at $-4$ ($-3.5$) PN than the double PSR. At $-4$PN order which corresponds to corrections due to e.g.~time variation in the gravitational constant $G$, the BH-PSR bounds are stronger than the double PSR one by almost two orders of magnitude. 
}
\end{center}
\end{figure}

We now give a brief summary of this paper.
In examining the prospects of bounding non-GR theories in a BH-PSR binary, it is important to compare to the existing method of binary PSR measurements. Figure~\ref{fig:gammaratio} presents the upper bound on the fractional non-GR correction $\gamma$ to the orbital decay rate with BH-PSRs relative to those with the double PSR, as a function of the entering PN order correction. If the ratio is below unity, BH-PSR bounds are stronger than the double PSR ones. Observe that the former can be stronger by orders of magnitude than the latter for negative PN corrections.

\begin{figure}[!h]
\begin{center}
\includegraphics[width=\linewidth]{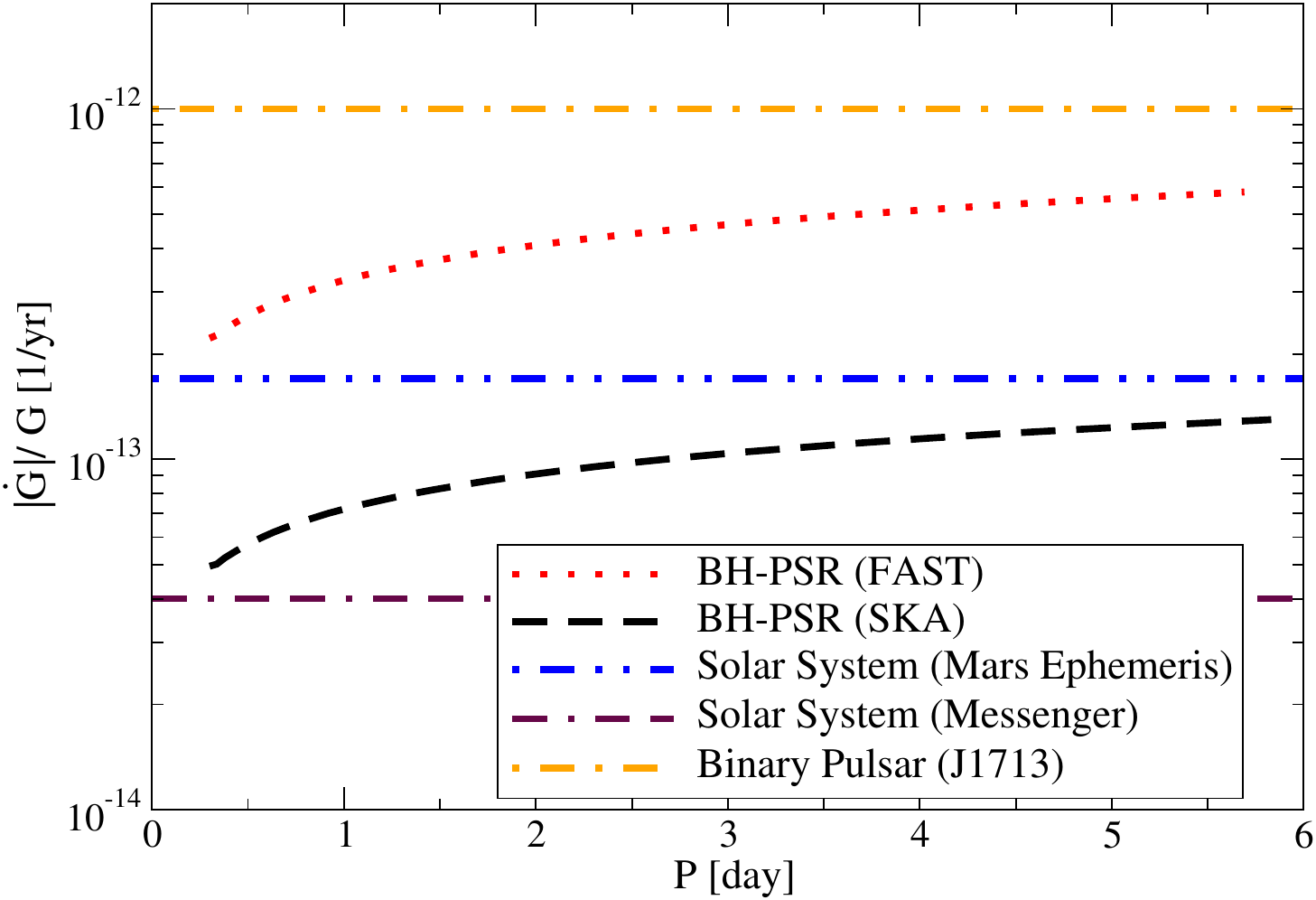}
\caption{\label{fig:VaryingG} Projected bounds on the time variation of the gravitational constant $G$ over the static gravitational constant as a function of the orbital period. The bound is shown for BH-PSR systems with FAST (red dotted) and SKA (black dashed). We also present two solar system bounds with The NASA Messenger (purple dotted-dashed line) \cite{2018NatCo...9..289G} and the Mars ephemeris (blue dotted-dashed line)~\cite{Will:2014kxa}, and the current strongest binary pulsar bound (dot dash orange) \cite{Zhu:2018etc,Zhu:2015mdo}. Notice that SKA can produce bounds that are comparable to or weaker than solar system ones, though the former are complementary to the latter as the two bounds probe different regime of gravity.
}
\end{center}
\end{figure}

Next, we apply such bounds on a generic non-GR parameter for the orbital decay rate to specific non-GR theories based on Table~\ref{tab:GammaValues}. For example, Fig.~\ref{fig:VaryingG} presents the upper bound on the time variation in $G$ as a function of the orbital period of BH-PSRs. Observe that BH-PSR observations with SKA are slightly weaker than the current strongest bounds from solar system experiments of NASA Messenger. However, since binary PSRs are sensitive to self gravity effects in the strong gravity regime, their tests complement a weak field test of $\dot G$ such as NASA Messenger. In the strong gravity regime, the time variation in $G$ can be magnified by a factor of 20 from that of weak field tests due to effects of the object's sensitivities \cite{WexBHP}. Binary PSR bounds also provide an independent test for non-GR effects. For other theories that we study in this paper, we find that BH-PSR $\dot P$ bounds are weaker than those obtained from NS-PSR or PSR-WD observations. 

Regarding bounds on quadratic gravity via BH quadrupole moment measurements, we find that bounds on dCS gravity can be improved by six to seven orders of magnitude for stellar-mass BH-PSR binaries. We also investigate bounding dCS gravity with a PSR orbiting Sgr~A$^*$, but we show that this system cannot reach a tight enough bound to satisfy the small coupling approximation. On the other hand, BH-PSR bounds on EdGB gravity are weaker than the current bounds from e.g.~BH-low-mass X-ray binaries (LMXB)~\cite{Yagi:2012gp} by an order of magnitude.

The rest of the paper is organized as follows. 
In Sec.~\ref{sec:Pdot}, we focus on orbital decay rate measurements. After discussing a generic formalism for describing non-GR corrections to the orbital decay rate and its relation to the PPE formalism in Sec.~\ref{sec:Pdot-general}, we study BH-PSR bounds in various modified theories of gravity in Sec.~\ref{sec:Pdot-example}.
In Sec.~\ref{sec:QuadrupoleMoment}, we focus on BH quadrupole moment measurements. After reviewing such measurements for BH-PSRs in Sec.~\ref{sec:Q-measurement}, we study bounds on two kinds of quadratic gravity in Sec.~\ref{sec:Q-example}.
We conclude in Sec.~\ref{sec:conclusion} and give possible avenues for future work.
We use the geometric units of $c=1$ and $G=1$ throughout the paper unless otherwise stated.

\section{Bounds from Orbital Decay Measurement}
\label{sec:Pdot}

Let us first focus on probing gravity with the measurement of the orbital decay rate $\dot P$ for BH-PSR binaries. We will first explain our generic formalism and show mapping between generic non-GR parameters entering in the orbital decay rate ($\gamma$ and $n$ mentioned earlier) to the theoretical parameters in example non-GR theories (column 5 of Table~\ref{tab:GammaValues}). We will next show the relation between such generic formalism with $\dot P$ to the PPE formalism~\cite{Yunes:2009ke}, which is a generic formalism to test strong-field gravity with GWs from compact binary mergers. We will then use the estimated measurement accuracy of $\dot P$ for BH-PSRs with FAST and SKA in~\cite{WexBHP} and derive projected bounds on generic modifications to $\dot P$. We will finally map these generic bounds to example non-GR theories.

\subsection{Non-GR Corrections to the Orbital Period Decay Rate}
\label{sec:Pdot-general}

\subsubsection{Formalism}

We will begin by considering the following generic non-GR modifications to $\dot P$~\cite{Yunes:2010qb}:
\begin{equation}
\label{eq:generic-Pdot}
\frac{\dot{P}}{P} = \frac{\dot{P}}{P}\biggr \rvert_\GR \Big(1 + \gamma \, v^{2 n}\Big)\,.
\end{equation}
Here $P$ is the orbital period, $v$ is the relative velocity of two compact objects in a binary, while the subscript ``GR'' means the quantity is evaluated in GR. $\dot P/P|_{\GR}$ is given by~\cite{Maggiore_GWBook}
\begin{equation}
\label{eq:Pdot-GR}
\frac{\dot{P}}{P} \Big\rvert_\GR = -\frac{96}{5} G^{5/3} \mu M^{2/3} \left( \frac{P}{2 \pi}\right)^{-8/3} F_\GR(e) \; ,
\end{equation}
where $M$ and $\mu$ are the total mass and the reduced mass respectively and
\begin{equation}\label{eq:FGR}
F_{\GR}(e) \equiv \frac{1}{(1-e^2)^{7/2}}\left( 1+\frac{73}{24}e^2+\frac{37}{96}e^4\right) \; .
\end{equation}

Each modification to GR is parameterized by $\gamma$ and $n$. The former gives the overall magnitude of the correction, while the latter tells us how the correction depends on $v$. In terms of the PN order counting, a correction term proportional to $v^{2 n}$ means that it enters at $n$-PN order relative to GR. Such a PN order counting gives us insight on what types of binaries have more advantage on probing specific types of modifications. For example, a theory with a negative PN correction would have a more stringent bound from a system with a smaller velocity (or widely-separated orbit) and vice versa for a theory with a positive PN. A selected example of ($\gamma$,$n$) in non-GR theories are presented in Table~\ref{tab:GammaValues}.

\begin{figure}[!h]
\begin{center}
\includegraphics[width=\linewidth]{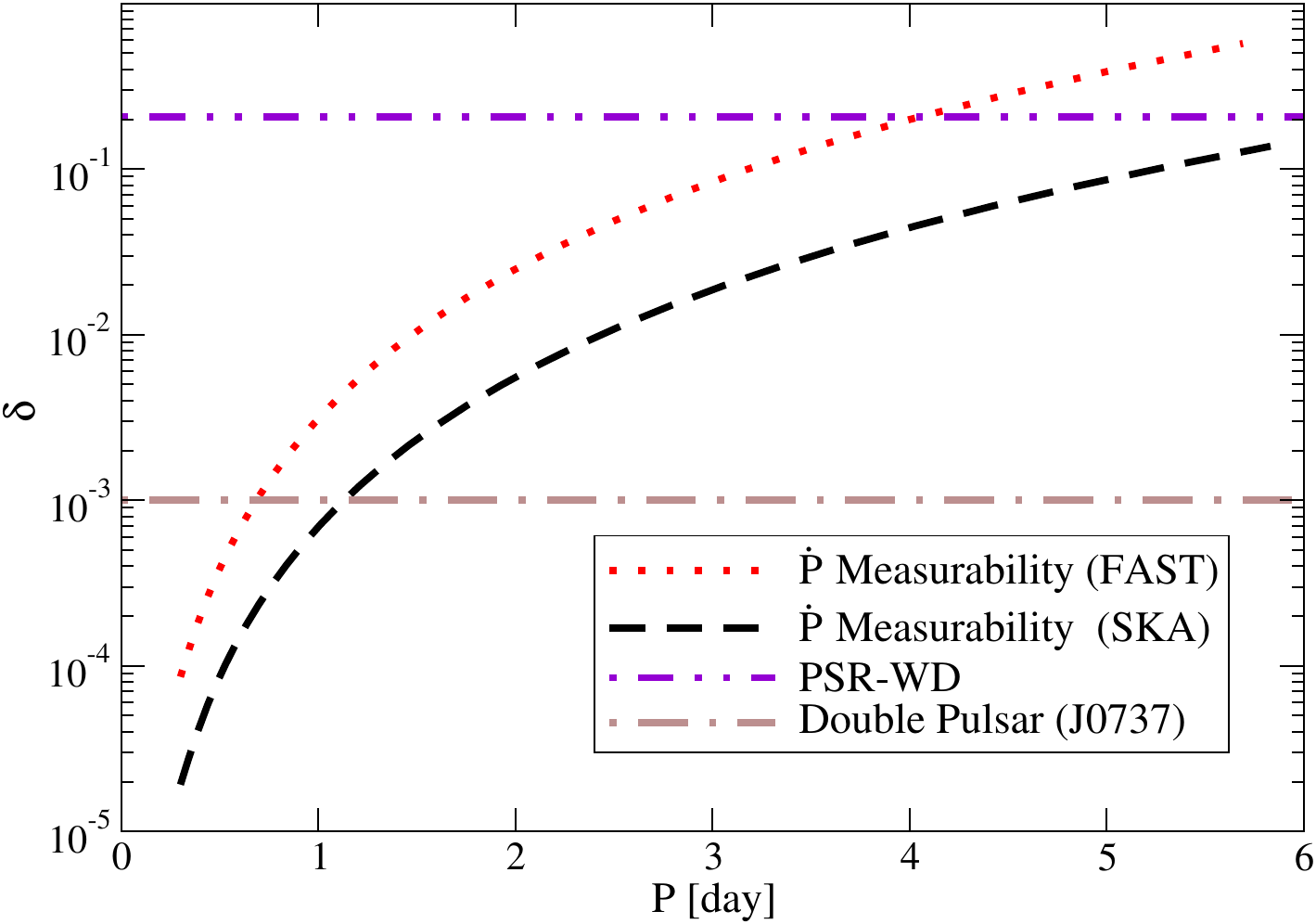}
\caption{\label{fig:PdotMeasurability} 
The measurability of the orbital decay rate $\dot P$ as a function of the orbital period $P$ with FAST (red dotted) and SKA (black dashed)~\cite{WexBHP}. The BH mass and the orbital eccentricity is assumed to be $10 M_\odot$ and $e=0.1$ respectively. For reference, we also present the measurability for a PSR-WD binary (purple double-dotted-dashed)~\cite{Freire:2012mg} and the double PSR (brown dotted-dashed)~\cite{Kramer:2016kwa}. 
}
\end{center}
\end{figure}

The projected measurement accuracy of $\dot P$ from BH-PSRs with FAST and SKA has been estimated in~\cite{WexBHP}, which we also present in Fig.~\ref{fig:PdotMeasurability}. Such a measurement accuracy is simulated for a system with a stellar-mass BH and a millisecond PSR as a function of orbital period. Reference~\cite{WexBHP} predicts the accuracy that FAST and SKA will achieve using a time of arrival (TOA) precision of 100 ns and 20 ns respectively. The simulation uses the following system parameters: eccentricity $e = 0.1$, 3 year observation period, $1.4 \Msolar$ PSR mass, $10 \Msolar$ BH mass, a $60^\circ$ inclination angle between the orbital angular momentum and the line of sight, and 4 hour weekly observations corresponding to 10 TOA. This simulation's results will be used in the remaining of Sec.~\ref{sec:Pdot} for BH-PSR orbital decay rate measurability. We can easily map these measurement accuracies to upper bounds on $\gamma$ for different PN correction terms as follows. Let us assume that $\dot P/P$ has been measured with a fractional error of $\delta$ as
\begin{equation}
\label{eq:Pdot-delta}
\left| \frac{\frac{\dot{P}}{P} - \frac{\dot{P}}{P} \rvert_\GR}{\frac{\dot{P}}{P} \rvert_\GR} \right| < \delta \; . 
\end{equation}
Combining this with Eq.~\eqref{eq:generic-Pdot}, one finds $\gamma$ can be constrained as 
\begin{equation}\label{eq:gammaboundvb}
|\gamma| < \frac{\delta}{v^{2n}} \;.
\end{equation}

\begin{figure}[!h]
\begin{center}
\includegraphics[width=\linewidth]{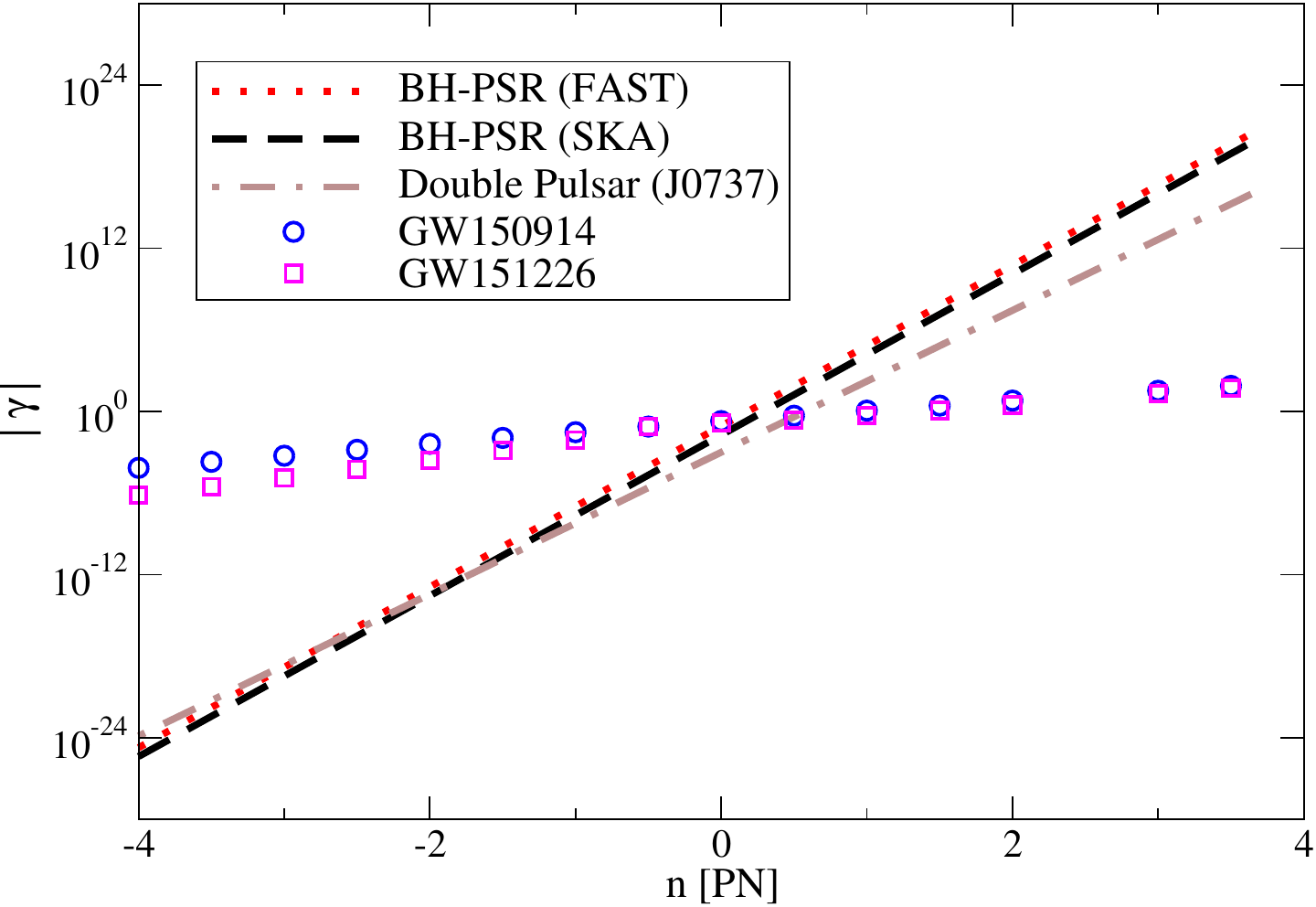}
\caption{\label{fig:gammapn} The upper bound on the fractional non-GR correction to the orbital decay rate $\gamma$ in Eq.~\eqref{eq:gammaboundvb} at each PN order for various astrophysical systems. We present projected bounds with a BH-PSR system using FAST (red dotted) and SKA (black dashed). We choose a 3 day orbital period for the BH-PSR binary. For comparison, we also show bounds from GW observations with GW150914 (blue circle) and GW151226 (magenta square)~\cite{Yunes:2016jcc}, together with those from the double PSR system (brown dotted-dashed)~\cite{Yunes:2010qb}. Observe that BH-PSR and double PSR bounds are much stronger for negative PN corrections, while GW bounds have more advantage on probing positive PN corrections.
}
\end{center}
\end{figure}

Figure~\ref{fig:gammapn} presents the projected upper bound on $\gamma$ as a function of at which PN order the correction enters, assuming that a BH-PSR with an orbital period of 3 days has been found by FAST or SKA. We choose this as an example system because it is an average case scenario for $\dot P$ measurability as shown in Fig.~\ref{fig:gammapn}. For comparison, we also present the bound from the double PSR~\cite{Yunes:2010qb}. Observe that the bounds from this BH-PSR share a similar trend compared to those from the double PSR, though the former can place more stringent bounds for negative PN corrections. This is because the relative velocity for the double PSR is $\sim 2 \times 10^{-3}$ while that for the BH-PSR assumed here is $\sim 1 \times 10^{-3}$ (in units of $c=1$). Although the difference is only a factor of 2, such a difference is enlarged if one considers negative PN corrections. For example, $-4$PN corrections become $2^8=256$ times larger for the BH-PSR than the double PSR case.

Figure~\ref{fig:gammaratio} explicitly shows the comparison on the bounds on $\gamma$ between the BH-PSR and the double PSR. If the ratio is below unity, the former is stronger than the latter. Observe that below approximately $-1$PN, the BH-PSR system can constrain non-GR corrections to $\dot P$ more strongly than the double PSR system. The former bound can be stronger than the latter one by many orders of magnitude for corrections at $-4$PN order.

\subsubsection{Relation to the PPE Formalism} \label{subsec:PPE}

Next, let us review the relation between the generic non-GR modification to $\dot P$ in Eq.~\eqref{eq:generic-Pdot} to that in gravitational waveforms from compact binary mergers. One example of the latter can be captured via the PPE formalism~\cite{Yunes:2009ke}. This is done by making an amplitude and phase correction to the gravitational waveform $\tilde h$ in the Fourier domain as~\cite{Yunes:2009ke} 
\begin{equation}
\tilde{h}(f) = \tilde{h}_\GR \left(1+\alpha \; u^a \right) e^{i \beta u^b}\,.
\end{equation}
Here $u \equiv 2 \pi \mathcal{M}/P$ where $\mathcal{M} \equiv M \eta^{3/5}$ is the chirp mass with $\eta \equiv m_p m_c/M^2$ representing the symmetric mass ratio. The PPE parameters, $\alpha$ and $\beta$, control the overall magnitude of non-GR corrections to the amplitude and phase respectively, while $a$ and $b$ show the dependence of such corrections to $u$. The GR waveform is recovered by setting $(\alpha,\beta)=(0,0)$. Such PPE formalism (or its modified version) has recently been applied to observed GW events~\cite{TheLIGOScientific:2016src,Yunes:2016jcc}.

One way to extract GW signals from the observed data is via matched filtering, where one cross-correlates the data against template waveforms. Because this analysis is more sensitive to phase corrections than amplitude corrections, many previous works including~\cite{TheLIGOScientific:2016src,Yunes:2016jcc} mentioned earlier only consider modifications to the phase. In this paper, we also follow this approach and set $\alpha=0$.

One can categorize non-GR modifications to compact binary evolution into two different classes, conservative and dissipative. The former corresponds to modifications to the binding energy (sum of the kinetic energy and gravitational potential energy) of the binary system, which also modifies Kepler's law. On the other hand, dissipative corrections modify the amount of energy being lost from binary systems due to emission of GWs and additional radiation (such as scalar radiation) if present in non-GR theories. 

When dissipative corrections dominate conservative ones, one can map PPE modifications in gravitational waveforms to $\dot P$ corrections by~\cite{Yunes:2010qb} as
\begin{equation}
\frac{\dot{P}}{P} = \frac{\dot{P}}{P}\biggr \rvert_\GR \left[1 + \frac{48}{5} b(b-1)\,\beta\, u^{b+\frac{5}{3}}\right]\,. \label{eq:OrbitalDecayRateBeta}
\end{equation}
Using $v = (2\pi M /P)^{1/3} = (u M/\mathcal{M} )^{1/3} = u^{1/3} \eta^{-1/5}$ and comparing the above equation with Eq.~\eqref{eq:generic-Pdot}, one finds 
\begin{equation}
\label{eq:gamma-beta}
\gamma = \frac{48}{5}\beta b(b-1) \eta^{\frac{3}{5}b+1}\;.
\end{equation}

Using Eq.~\eqref{eq:gamma-beta}, one can map bounds on $\beta$ from GW150914 and GW151226 in~\cite{Yunes:2016jcc} to those on $\gamma$. We present such results in Fig.~\ref{fig:gammapn}. Observe that BH-PSRs and NS-PSR/PSR-WD binaries have more advantage on constraining theories with negative PN corrections compared to GW observations~\cite{Yunes:2010qb}. We justify this statement for GW170817 by comparing its rough bounds in App.~\ref{ap:GW170817} to those of binary PSRs. This is because the compact objects in the former binaries move much slower than the latter binaries that are about to coalesce.

\subsection{Example Theories and Projected Bounds}
\label{sec:Pdot-example}

We now study projected BH-PSR bounds on specific example non-GR theories based on Fig.~\ref{fig:gammapn} and Table~\ref{tab:GammaValues}. Such BH-PSR bounds have already been estimated within the context of scalar-tensor theories~\cite{WexBHP,Wex_ScalarTensor}, EdGB gravity~\cite{EDGBBoundComparison}, a brane-world model~\cite{Simonetti:2010mk} and quantum gravity~\cite{Estes:2016wgv}. In this section, we study four different theories; theories with time-varying gravitational constant $G$, Lorentz-violating massive gravity, cubic Galileon theories and generic screened modified gravity. 

\subsubsection{Varying G Theories} \label{subsec:VaryingG}
The gravitational constant's value can be time dependent in many modified theories of gravity~\cite{Will:2014kxa}. This is the case when $G$ depends on the scalar field that is coupled to the metric (like scalar-tensor theories in the Jordan frame). Such time variation in $G$ can affect the orbital decay rate in two ways, conservative ($\dot P/P|_{C}$) and dissipative ($\dot P/P|_{D}$), as already mentioned earlier. In most literature, only the former is included (see e.g.~\cite{VaryingGNordtvedt,WexRadioPulsars}). Here we explicitly show why the latter is highly suppressed compared to the former.

Let us first look at dissipative corrections. Such corrections can be derived from $G$ dependence on $\dot P/P|_\GR$ in Eq.~\eqref{eq:Pdot-GR}, which is $\dot P/P|_\GR \propto G^{5/3}$. We now promote $G$ to include the time dependence. We do this by assuming the time variation is sufficiently small and Taylor expand $G$ about the initial observation time $t_0$ and keep up to linear order in $t$: $G(t) \approx G_0+\dot{G}(t-t_0)$. Then, one finds 
\begin{eqnarray}
 \frac{\dot{P}}{P}\propto G^{5/3} &=& [G_0+\dot{G}(t-t_0)]^{5/3} \nonumber \\
 & \approx & G_0^{5/3}\left(1+\frac{5}{3} \frac{\dot{G}}{G_0} (t-t_0)\right) \;.
\end{eqnarray}
Thus, the dissipative correction is given by
\begin{equation}
\frac{\dot{P}}{P} \biggr \rvert_{D} = \frac{5}{3} \frac{\dot{G}}{G_0}(t-t_0) \,\frac{\dot{P}}{P} \biggr \rvert_\GR\;.
\end{equation}
Hereafter we will drop the subscript ``0'' on $G_0$. 

Next, we look at conservative corrections, which are derived by taking the time derivative of the orbital period $P$, assuming that there is no gravitational radiation. The orbital period is given by~\cite{VaryingGNordtvedt}
\begin{equation}
P =\frac{1}{(1-e^2)^\frac{3}{2}} \frac{2 \pi l^3}{(G M)^2} \; ,
\end{equation}
where $l$ is the specific orbital angular momentum. Taking the time derivative, one finds 
\begin{equation}
\frac{\dot{P}}{P} \biggr \vert_C = -2\frac{\dot{G}}{G}-2 \frac{\dot{M}}{M} +3 \frac{\dot l}{l} \;.
\end{equation}
This equation can be reduced further to~\cite{VaryingGNordtvedt}
\begin{equation}
\label{eq:Gdot-cons}
\frac{\dot{P}}{P} \biggr \vert_C = -2 \frac{\dot{G}}{G}\left[1 - \left(1+\frac{m_c}{2M}\right)s_p-\left(1+ \frac{m_p}{2M}\right)s_c \right] \;,
\end{equation}
where $m_p$ and $m_c$ are PSR mass and companion mass respectively and the sensitivity is defined as
\begin{equation}\label{eq:sensitivity}
s_A = - \frac{\partial \ln m_A}{\partial \ln G} \; ,
\end{equation}
which measures how the mass depends on $G$.
In GR, $s_A=0$~\cite{WexRadioPulsars}. 

In order to estimate the relative strength of the conservative and dissipative corrections, let us consider Damour-Esposito-Far\` ese scalar-tensor theories as an example. We choose $m_p=1.4M_\odot$ and $m_c=10M_\odot$. In this theory, BHs have $s_c = 0.5$, while $s_p$ for PSRs depend on the underlying equation of state, though typically one finds $s_p \sim 0.15$ for $m_p=1.4M_\odot$ (see Fig.~20 of~\cite{WexRadioPulsars}). Assuming further $P = 0.1 $ day and $t-t_0 = 5$ years, the ratio of the two corrections becomes 
\begin{equation}
 \frac{\dot{P}}{P} \biggr \rvert_{D} / \frac{\dot{P}}{P} \biggr \vert_C \approx 10^{-7} \;.
\end{equation}
Thus dissipative corrections are highly suppressed relative to conservative corrections and can be ignored. The suppression is due to the fact that the radiation reaction timescale for BH-PSRs and other binary PSRs is very large compared to the observational time. This is not the case for coalescing compact binaries and dissipative corrections are important for GW observations~\cite{Yunes:2009bv,Tahura}.

Having these pieces of information in hand, we can now estimate future bounds on time variation in $G$ from BH-PSR binaries. Using Eqs.~\eqref{eq:generic-Pdot} and~\eqref{eq:gammaboundvb}, the non-GR parameters $(\gamma,n)$ for varying $G$ theories are given by the expression in~Table~\ref{tab:GammaValues}. Notice that $\gamma$ is proportional to $\dot G$ and the correction enters at $-4$PN order. Using further Eq.~\eqref{eq:Gdot-cons}, the upper bound on $\dot G$ from the uncertainty in the orbital decay rate measurement is given by
\begin{equation}
\frac{|\dot{G}|}{G} < -\frac{1}{2} \frac{\dot{P}}{P} \biggr \vert_\GR \frac{\delta}{1 - \left(1+\frac{m_c}{2M}\right)s_p-\left(1+ \frac{m_p}{2M}\right)s_c} \;.\label{eq:Gdotbound}
\end{equation}
Figure~\ref{fig:VaryingG} presents projected upper bounds on $\dot G$ from BH-PSRs with FAST and SKA. Here we assume scalar-tensor theories for calculating sensitivities of BHs and PSRs. For comparison, we also show current bounds from solar system experiments~\cite{Will:2014kxa}. Observe that the projected BH-PSR bounds are slightly weaker than solar system experiments from NASA Messenger \cite{2018NatCo...9..289G,2015CeMDA.123..325F}. We do not show bounds from NS-PSR/PSR-WD binaries and recent GW observations as they are much weaker than BH-PSR bounds~\cite{Will:2014kxa,Yunes:2016jcc} (see App.~\ref{ap:GW170817} for a rough bound from GW170817). Of course, NS-PSR and PSR-WD tests will also strengthen in the future by the same improved radio telescopes.

A BH-PSR constraint on $\dot G$ is useful to include with stronger solar system measurements. Solar system experiments, such as NASA Messenger, measure time variation in $G$ differently than strongly self gravitating bodies. First, the measurement is of $(\partial_t(G \Msolar))/(G \Msolar) $ instead of $\dot G / G$, so uncertainty in the solar mass and its time derivative couple into the bound in time variation in G. More importantly, binary PSR measurements capture new effects not present in solar system experiments. Strong field effects can enhance $\dot G$ compared to that in the weak field. For example, Fig.~8 of Ref.~\cite{WexRadioPulsars} shows the $\dot G$ effect has an enhancement factor of over an order of magnitude in the strong field in certain scalar-tensor theories (which is not reflected in Fig.~\ref{fig:VaryingG}). Thus, there is a possibility that future strong-field observations of BH/PSRs can detect non-vanishing $\dot G$ and yet still satisfying the solar system bounds, though the detailed calculation will be left for future work.

\subsubsection{Lorentz-violating Massive Gravity} \label{subsec:FierzPauli}

Next, let us study Lorentz-violating massive gravity, which is an extension to GR where the graviton is assumed to have a non-vanishing mass $m_g$ (see e.g.~\cite{Rubakov:2008nh,Hinterbichler:2011tt,deRham:2014zqa} for reviews on massive gravity). Historically, Fierz and Pauli~\cite{Fierz:1939ix} constructed a Lorentz-invariant massive gravity about Minkowski background. A linearized version of such a theory does not reduce to that of GR in the limit $m_g \to 0$ (known as the van Dam-Veltman-Zakharov discontinuity), though GR can be correctly recovered once one takes non-linear effects into account (the Vainshtein mechanism)~\cite{1972PhLB...39..393V}. 

The Lorentz-violating version of massive gravity that we consider here was studied by~\cite{Finn:2001qi} within the context of binary PSRs. The action is modified from Fierz-Pauli massive gravity with two important properties. First, the $m_g \rightarrow 0$ limit of the linearized Lorentz-violating massive gravity recovers that of GR. Second, the field equations for metric perturbations $h_{\mu\nu}$ in Lorentz gauge to its linear order are described simply by
\begin{equation}
(\square -\bar m_g^2)h_{\mu \nu} = -16\pi T_{\mu \nu} \;,\label{eq:FPFieldEquationPerturbation}
\end{equation}
where $\bar m_g \equiv m_g/\hbar$ and $T_{\mu \nu}$ is the matter stress energy tensor that is independent of the metric perturbation~\cite{Finn:2001qi}. 

Corrections to $\dot P$ in Lorentz-violating massive gravity come from 
the dissipative sector, namely modifications to GW emission. One can compute the amount of GWs being emitted from compact binaries in this theory by solving Eq.~\eqref{eq:FPFieldEquationPerturbation} using a Green's function. One can then calculate gravitational luminosity $L$, whose fractional difference from the GR expression is related to that for $\dot P$ by 
\begin{equation}\label{eq:pdotGW}
\frac{\dot{P} - \dot{P}_\GR}{\dot{P}_\GR} = \frac{L- L_\GR}{L_\GR} \;.
\end{equation}
One can then read off $(\gamma,n)$ for Lorentz-violating massive gravity, as shown in Table~\ref{tab:GammaValues}.
Notice that $\gamma$ is proportional to $m_g^2$ as the mass of the graviton enters as $m_g^2$ in the modified wave equation in Eq.~\eqref{eq:FPFieldEquationPerturbation}, and the correction to $\dot P$ enters at $-3$PN order. 
 Combining the expression for $\gamma$ with Eq.~\eqref{eq:Pdot-delta}, Finn and Sutton~\cite{Finn:2001qi} derived bounds on the mass of the graviton as 
\begin{equation}\label{eq:massfp}
m_g^2 \leq \frac{24}{5} (1-e^2)^{1/2} F_\GR(e) \left(\frac{2 \pi \hbar}{c^2 P}\right)^2 \delta \; .
\end{equation}

\begin{figure}[!h]
\begin{center}
\includegraphics[width=\linewidth]{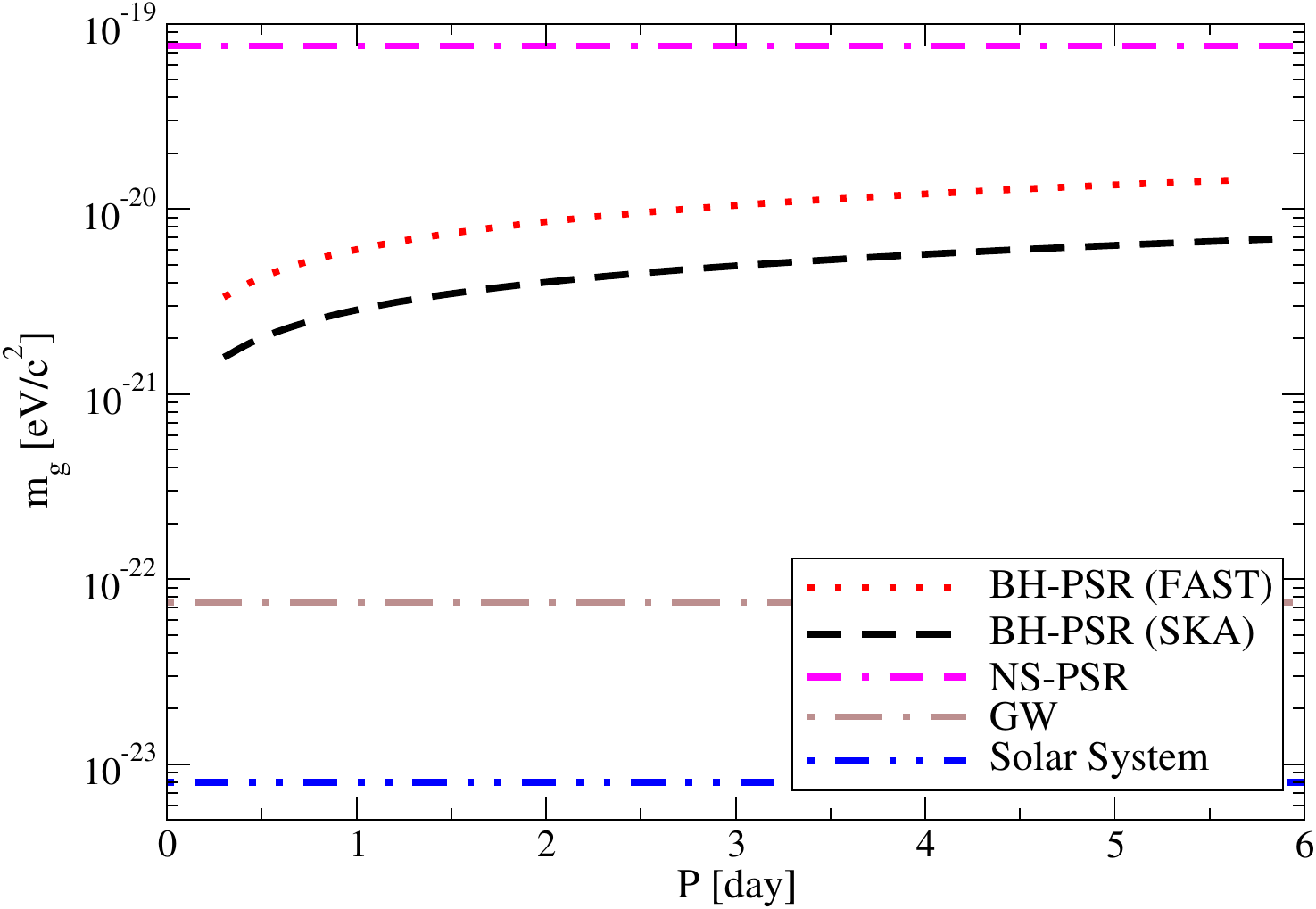}
\caption{\label{fig:fpmassivegravity} Bounds on the mass of the graviton in Lorentz-violating massive gravity calculated from Eq.~\eqref{eq:massfp}. We present projected bounds with BH-PSRs using FAST (red dotted) and SKA (black dashed) as a function of its orbital period. For comparison, we also present the strongest bound from a combination of PSR B1913+16 and B1534+12 (magenta dotted-double-dashed) \cite{Finn:2001qi} and GWs (brown dotted-dashed)~\cite{Abbott:2017vtc}. Notice that the BH-PSR system can place a stronger bound than the binary PSR one but the former is still weaker than the GW ones. The strongest current bound comes from solar system measurements of the perihelion advance of Mars and Saturn (blue double-dotted-dashed)~\cite{GravitonSSBound}. 
} 
\end{center}
\end{figure}

Figure~\ref{fig:fpmassivegravity} presents the projected upper bounds on $m_g$ from BH-PSRs with FAST and SKA. For comparison, we also present bounds from solar system experiments~\cite{GravitonSSBound} (which updates the previous bounds~\cite{Talmadge:1988qz} by nearly two orders of magnitude), NS-PSR observations of PSR B1913+16 and B1534+12~\cite{Finn:2001qi} and recent GW events~\cite{Abbott:2017vtc} (see e.g. \cite{Desai:2017dwg} for a model-dependent bound on mass of the graviton from galaxy cluster observations). Observe that the projected BH-PSR bounds are stronger than the NS-PSR ones by more than an order of magnitude. As mentioned in Sec.~\ref{sec:intro}, this is because the relative velocity of two compact objects in a binary is smaller for BH-PSRs than for NS-PSRs. On the other hand, such BH-PSR bounds are not as stringent as the solar system or GW bounds. Gravitational wave bounds are much stronger because they probe corrections to the modified dispersion relation of the graviton, which accumulates over distance during GW propagation.

\subsubsection{Cubic Galileon Massive Gravity}\label{subsec:CubicGalileon}

We now study another type of massive gravity. Lorentz-violating massive gravity studied in the previous subsection captures a generic massive gravity modification in the wave equation or the dispersion relation of the graviton given by Eq.~\eqref{eq:FPFieldEquationPerturbation}. Another important aspect of massive gravity is the screening effect called the Vainshtein mechanism~\cite{1972PhLB...39..393V}, where the scalar degrees of freedom (that arise from additional graviton helicity states in massive gravity) become strongly-coupled within the Vainshtein radius, which suppresses deviations away from GR. Many of the features of this mechanism can be generically captured via the Galileon models~\cite{Nicolis:2008in}. Indeed, the generic Galileon arises from the ghost-free massive gravity~\cite{deRham:2010kj} in a certain limit~\cite{deRham:2010ik,deRham:2010gu}. Following~\cite{MassiveGravityCG}, we consider one of the simplest types of such models here, namely the cubic Galileon model. Such a model is still allowed from GW170817, while many of the other Galileon models have been ruled out~\cite{Baker:2017hug,Ezquiaga:2017ekz}. Galileon models are also motivated from explaining the current accelerating expansion of our universe. In this subsection, we use the units $c=1$, $G=1$ and $\hbar=1$.

The Galileon radiation can be found by varying the action in the decoupling limit of infinite Planck mass~\cite{MassiveGravityCG}, in which one can neglect the self-interactions of the helicity-two graviton. In this theory, there is not only quadrupolar Galileon radiation but also lower order contributions, namely monopolar and dipolar radiation. However, the former has the largest effect in terms of PN expansion~\cite{MassiveGravityCG}. Thus, in this paper we focus on corrections to $\dot P$ due to the quadrupolar Galileon radiation. The correction to the GR GW luminosity due to such radiation is given by~\cite{MassiveGravityCG}
\begin{equation}
L - L_\GR = \frac{5 \lambda^2}{32} \frac{(\Omega_P a)^3}{(\Omega_P r_\star)^\frac{3}{2}}\frac{M_\Q^2}{M_\PL^2}\Omega_P^2 F_\CG(e)\;,\label{eq:CGpower}
\end{equation}
where $\Omega_P$ is the angular orbital frequency, $a$ is the semi-major axis, $M_\PL$ is the Planck mass and $M_\Q$ is defined as
\begin{equation}\label{eq:MQdef}
M_\Q = \frac{m_p m_c (\sqrt[]{m_p}+\sqrt[]{m_c})}{M^\frac{3}{2}} \;.
\end{equation}
$\lambda$ is a numerical factor given by\footnote{Our normalization of $\lambda$ in Eq.~\eqref{eq:lambdadef} and $I_k^Q$ in Eq.~\eqref{eq:CGI_n^Q} is slightly different from that in~\cite{MassiveGravityCG}. Our normalization is chosen such that $F_\CG(0)=1$.}
\begin{equation}\label{eq:lambdadef}
\lambda = \frac{3^\frac{9}{8}\pi^\frac{1}{4}}{2^\frac{5}{2}\Gamma(9/4)} \;,
\end{equation}
while $r_\star$ is the Vainshtein radius inside which nonlinear effects become important and is given by
\begin{equation}
r_\star = \left( \frac{M}{16 m_g^2 M_\PL^2}\right)^\frac{1}{3} \;.
\end{equation}
$F_\CG(e)$ is defined as
\begin{equation}\label{eq:FCGdef}
F_\CG(e)=\sum_{k=0}^\infty|I_k^Q(e)|^2 \; ,
\end{equation}
with $I_k^Q$ given by
\begin{equation}\label{eq:CGI_n^Q}
I_k^Q=\frac{(1-e^2)^\frac{3}{2}}{2\pi} \left( \frac{k}{2}\right)^\frac{7}{4}\int_0^{2\pi}\frac{\exp\left[i(2-k)x\right]}{(1+e\cos{x})^\frac{3}{2}}dx\;.
\end{equation}

For the monopole and dipole radiation that we do not consider in our paper, an analytic solution to similar integrals as in Eq.~\eqref{eq:CGI_n^Q} can be found~\cite{MassiveGravityCG}. For completeness, we correct typos in the dipole analytic expression in~\cite{MassiveGravityCG} in App.~\ref{ap:DipoleCGCorrection}. On the other hand, one needs to evaluate Eq.~\eqref{eq:CGI_n^Q} numerically for the quadrupole radiation since the fractional power in the denominator prevents integration by the same method. Although the first 15 terms in $k$ give the dominant contribution, we keep up to $k=30$ to reduce the error from the infinite summation result.

Next, let us derive the upper bound on $m_g$ in terms of the fractional measurement accuracy of the orbital decay rate $\delta$. First, using Eqs.~\eqref{eq:generic-Pdot},~\eqref{eq:pdotGW} and~\eqref{eq:CGpower}, one can derive $(\gamma,n)$ for cubic Galileon theories that are given in Table~\ref{tab:GammaValues}. Notice that $\gamma$ is proportional to $m_g$ (unlike the $m_g^2$ scaling for Lorentz-violating massive gravity) and the correction enters at $-11/4 (= -2.75)$PN order (which is close to $-3$PN order in Lorentz-violating massive gravity). One can then use Eq.~\eqref{eq:gammaboundvb} to find
\begin{equation}
m_g \leq \frac{2^3}{5 \lambda^2}\frac{1}{F_\CG(e)}\frac{M^\frac{1}{2}M_\PL}{ M_\Q^2}\frac{1}{\Omega_P^\frac{1}{2} (\Omega_P a)^3} L_\GR \delta\; .\label{eq:CGmassbound}
\end{equation}

\begin{figure}[!h]
\begin{center}
\includegraphics[width=\linewidth]{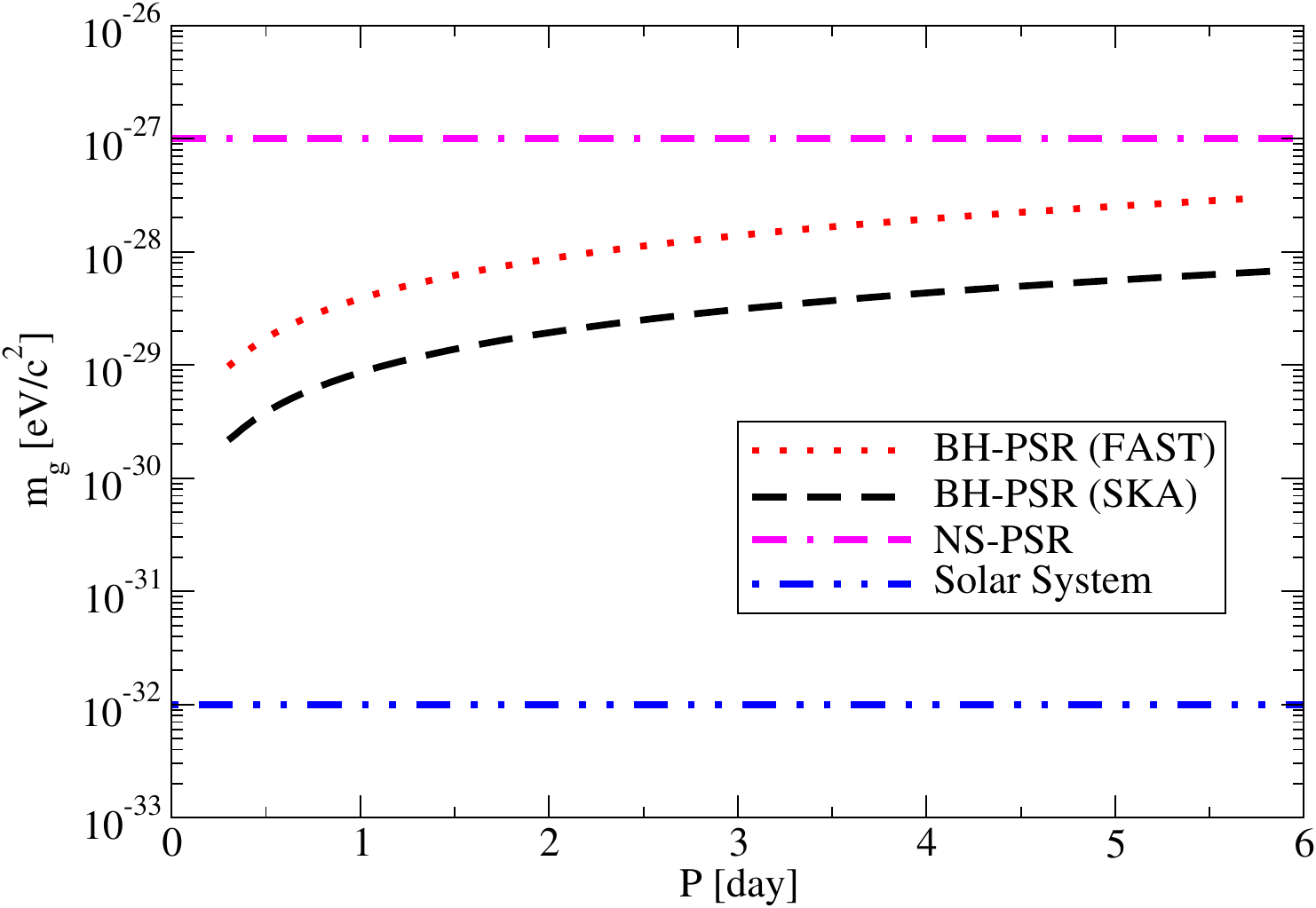}
\caption{\label{fig:CG} Similar to Fig.~\ref{fig:fpmassivegravity} but for cubic Galileon type massive gravity as calculated from Eq.~\eqref{eq:CGmassbound}. Observe that BH-PSRs yield stronger bounds than NS-PSR systems (magenta dotted-dashed)~\cite{MassiveGravityCG}, but weaker bounds than solar system bounds (blue double-dotted-dashed)~\cite{Dvali:2002vf}.
}
\end{center}
\end{figure}

Figure~\ref{fig:CG} presents projected upper bounds on the mass of the graviton in cubic Galileon massive gravity from BH-PSRs with FAST and SKA. We also present current bounds from NS-PSR observations~\cite{MassiveGravityCG} and solar system experiments~\cite{Dvali:2002vf}. Observe that the BH-PSR bounds are stronger (weaker) than NS-PSR (solar system) ones, which is similar to Lorentz-violating massive gravity case in Fig.~\ref{fig:fpmassivegravity}. Observe also that the bounds in cubic Galileon massive gravity are stronger than those in Lorentz-violating massive gravity. This is because the upper bound on $m_g$ scales linearly with $\delta$ for the former (see Eq.~\eqref{eq:CGmassbound}) since the Vainshtein suppression to the quadrupolar gravitational radiation is linearly proportional to $m_g$, while such a bound scales with $\delta^{1/2}$ for the latter (see Eq.~\eqref{eq:massfp}) since the correction to the graviton dispersion relation scales with $m_g^2$.

\subsubsection{General Screened Modified Gravity}\label{subsec:SMG}

General screened modified gravity (SMG) is a scalar modification to GR with a fifth force and screening mechanism. The scalar field induces non-GR effects on cosmological scale that can explain current accelerating expansion of our universe without introducing dark energy. In general, scalar field also induces a fifth force that confronts solar system tests of gravity. One way to cure this problem is to introduce a screening mechanism that ensures one recovers GR within the solar system~\cite{Jain:2010ka,Joyce:2014kja,Koyama:2015vza}. In this section, we follow~\cite{Zhang:2017srh,Zhang:2018dxi,Liu:2018sia} and consider a generic screening effect (including chameleon~\cite{Khoury:2003rn,Khoury:2003aq,Gubser:2004uf}, symmetron~\cite{Hinterbichler:2010es,Hinterbichler:2011ca,Davis:2011pj} and dilaton~\cite{Damour:1994zq,Damour:1994ya,Brax:2010gi} mechanisms), where the scalar field acquires a mass in high density regimes and the mediation of additional degrees of freedom is suppressed\footnote{Notice that the Vainshtein mechanism~\cite{1972PhLB...39..393V} is another type of screening effect which we treated separately in Sec.~\ref{subsec:CubicGalileon}.}. Bounds on SMG have been derived from PSR-WD binaries \cite{Zhang:2017srh,Zhang:2018dxi}. Here, we study the prospect of probing SMG with future BH-PSR observations.

The orbital decay rate is modified by the scalar field in BH-PSR binaries through scalar dipole radiation. A generic expression for the orbital decay rate can be found in Eq.~(5) of \cite{Zhang:2018dxi}. Keeping only to leading correction in terms of PN expansion, one finds 
\begin{eqnarray}\label{eq:Screen_Grav_ODR}
\frac{\dot P}{P} &=& - \frac{m_p m_c}{M^{1/3}}\left(\frac{P}{2\pi} \right)^{-8/3}\left[\frac{96}{5} F_\GR(e) \right. \nonumber\\
&& \left. +\frac{F_\SMG(e)}{2}(\epsilon_p-\epsilon_c)^2\left(\frac{P}{2\pi M} \right)^{2/3}\right]\; ,\end{eqnarray}
with $F_\SMG(e)$ defined as
\begin{equation}
\label{eq:F_SMG}
F_\SMG(e) \equiv \frac{2+e^2}{2(1-e^2)^{5/2}}\; ,
\end{equation}
and the screening parameter (similar to the scalar charge) of the $A$th body is given by
\begin{equation}\label{eq:ScreenScalarCharge}
\epsilon_A = \frac{ \phi_{\VEV}-\phi_A}{M_\PL \Phi_A} \;.
\end{equation}
$\Phi_A = m_A / R_A$ is the compactness of the $A$th body with $R_A$ denoting its radius, $\phi_A$ is the value of the scalar field at the minimum of the scalar field potential inside the star, and $\phi_{\VEV}$ is the vacuum expectation value of the scalar field. In Eq.~\eqref{eq:ScreenScalarCharge}, $\phi_{\VEV}>>\phi_A$. Thus, following~\cite{Zhang:2017srh,Zhang:2018dxi,Liu:2018sia}, we will neglect $\phi_A$ for the remainder of the paper. The screening mechanism is apparent in Eq.~\eqref{eq:ScreenScalarCharge} due to the inverse proportionality of the compactness. Therefore, objects with smaller compactness dominate the screening parameter and thus contribute the majority of the radiation in the scalar field. For BHs, screening parameters vanish~\cite{Zhang:2017srh} due to the no-hair theorem in scalar-tensor theories~\cite{Hawking:1972qk,Bekenstein:1995un,Sotiriou:2011dz}. 

We now derive projected bounds on SMG with future BH-PSRs observations. First, one can easily read off $\gamma$ and $n$ in Eq.~\eqref{eq:generic-Pdot} from Eq.~\eqref{eq:ScreenScalarCharge} for SMG, as shown in Table~\ref{tab:GammaValues}. Notice that $\gamma$ is proportional to $(\epsilon_p - \epsilon_c)^2$ and the correction enters at $-1$PN order. Such a structure is similar to that in scalar-tensor theories and EdGB gravity. One can next use Eq.~\eqref{eq:gammaboundvb} to find the following expression for the upper bound on $\phi_\VEV$, 
\begin{equation}\label{eq:UpperBoundVEV}
\left|\frac{\phi_\VEV}{M_\PL}\right|\leq \frac{m_p}{R_p}\left(\frac{2\pi M }{P}\right)^{1/3}\left[ \frac{192}{5} \frac{F_\GR(e)}{F_\SMG(e)}\delta\right]^{1/2} \; .
\end{equation}

\begin{figure}[!h]
\begin{center}
\includegraphics[width=\linewidth]{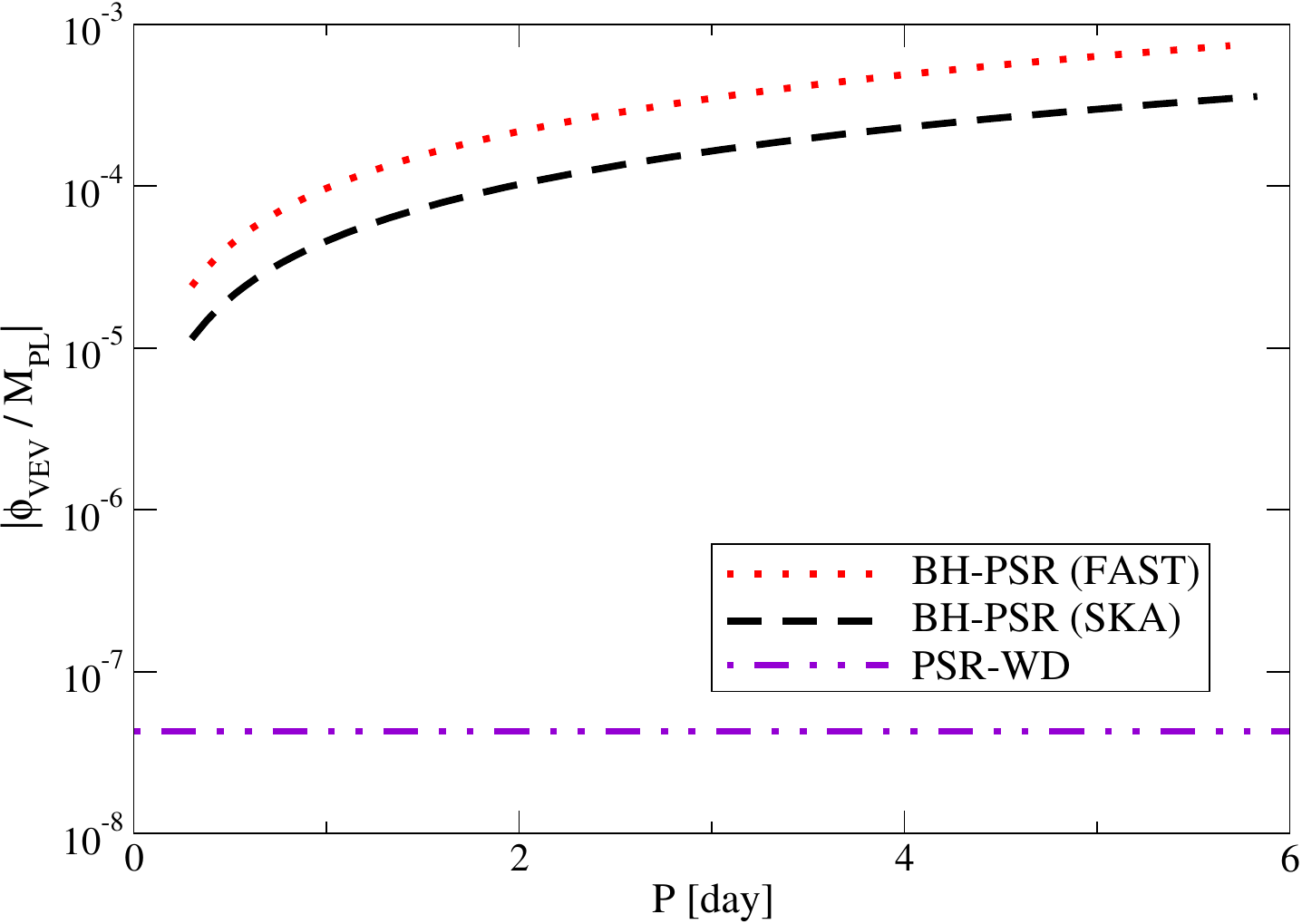}
\caption{\label{fig:Screened_Grav}The upper bound on vacuum expectation value of the scalar field per Planck mass in SMG as a function of orbital period by a measurement of orbital decay rate for BH-PSR binaries using FAST (red dotted) and SKA (black dashed) calculated from Eq.~\eqref{eq:UpperBoundVEV}. We also show the bound obtained from PSR-WD binaries (purple double-dotted-dash) in~\cite{Zhang:2018dxi}. Observe that using a PSR-WD has strong advantages over a BH-PSR since the WD compactness is approximately $10^4$ times larger than that of a PSR and the screening effect is less efficient. 
}
\end{center}
\end{figure}

Figure~\ref{fig:Screened_Grav} presents projected upper bounds on $\phi_\VEV/M_\PL$ as a function of the orbital period of BH-PSRs using FAST and SKA. The fiducial value of the PSR radius has been chosen as $R_{p} = 12 $ km. For reference, we also show the bound from PSR-WD binaries in~\cite{Zhang:2018dxi}. As expected, the latter is much stronger than the former, as non-GR effects in PSRs and BHs are highly suppressed due to either the screening effect or the no-hair theorem as opposed to WDs.

\section{Bounds from BH Quadrupole Moment Measurement}\label{sec:QuadrupoleMoment}

In this section, we will study an alternative way to probe gravity with BH-PSRs, namely via measurements of the BH quadrupole moment. The orbital decay rate used in the previous section has advantage on probing theories which generate negative PN corrections. This is the case for theories studied in Secs.~\ref{sec:Pdot-example}.
On the other hand, some theories give rise to modifications only at positive PN orders. If the BH quadrupole moment is different from that for a Kerr BH, measuring this quantity has more advantage on constraining such theories as we explain in more detail below.

\subsection{Quadrupole Moment Measurement with BH-PSRs} 
\label{sec:Q-measurement}

Let us first briefly review how one can measure the BH quadrupole moment in BH-PSR binaries. 
A non-vanishing quadrupole moment $Q$ of a BH produces a periodic perturbation of the PSR's orbit from one periastron passage to the next~\cite{Wex:1998wt}.
One of the most precise ways of measuring the quadrupole moment is through the Roemer time delay~\cite{QuadrupoleBoundFunctionofPb,WexBHP}, which is an modulation in travel time for light due to the PSR's orbit. The PSR's orbit around the binary's center of mass changes the distance between the PSR and earth. The Roemer time delay is thus defined to be the time for light to travel from closest and furthest locations of the PSR to earth (with the signal being barycentered so that modulation in the earth's orbit is removed). Defining a dimensionless parameter $\epsilon = -3 Q/a^2(1-e^2)^2$, the Roemer delay can be approximated as
\begin{equation}
\Delta_R =\Delta_R^{(0)}+\Delta_R^{(1)}+\mathcal{O}(\epsilon^2) \; ,
\end{equation}
where $\Delta_R^{(0)}$ is the contribution without the quadrupole moment at $\mathcal{O}(\epsilon^0)$, while $\Delta_R^{(1)}$ is the first correction due to the non-vanishing quadrupole moment at $\mathcal{O}(\epsilon)$. 

Let us next compare the Roemer time delay against orbital decay in terms of measuring the BH quadrupole moment. The former can directly measure the effect of quadrupole moment entering at 2PN order relative to the Newtonian Kepler motion in the orbital evolution. The quadrupole moment also affects the orbital decay rate at 2PN order relative to its leading effect. However, the leading radiation reaction effect (backreaction of GW emission onto the orbit) only enters from 2.5PN order in the orbital evolution, and thus, the effect of quadrupole moment to the orbital decay affects the orbital evolution at 4.5PN order and higher. This is why the Roemer time delay has more advantage on measuring the BH quadrupole moment compared to the orbital decay measurement.

\begin{figure}[!h]
\begin{center}
\includegraphics[width=\linewidth]{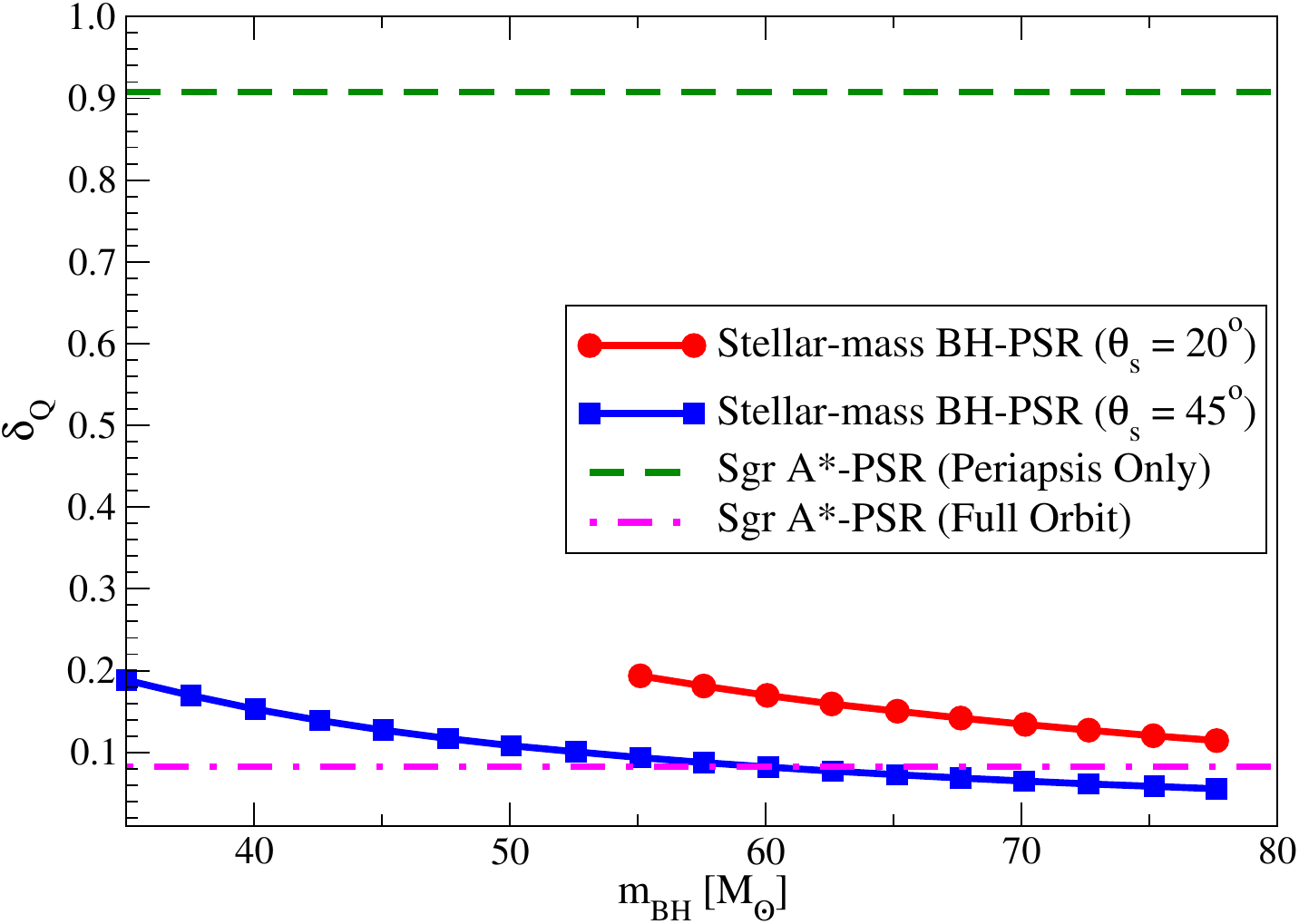}
\caption{\label{fig:Quad_Mom_Measure}
The fractional measurability of the BH quadrupole moment with stellar-mass BH-PSRs as a function of the BH mass for two different spin inclination angles $\theta_s$~\cite{WexBHP}. This assumes a 20 year observation length with SKA, the dimensionless BH spin of $\chi = 1$ and the orbital eccentricity of $e=0.5$. The time to coalesce has been fixed to 10 Myr, which corresponds to the orbital period of 0.16 days ($m_\BH = 30M_\odot$) to 0.21 days ($m_\BH = 80M_\odot$). We also plot the simulated measurability of the quadrupole moment of Sgr A* by pulsar timing from \cite{QuadrupoleMeasureabilitySgrA_PsaltisWex}. Periapsis only contains measurements by measuring three periastron passages, while full orbit considers the entire orbit in the measurement. 
}
\end{center}
\end{figure}

Two possible types of BH-PSR binaries exist for probing gravity via the BH quadrupole moment measurements: stellar-mass BH-PSRs and supermassive BH-PSRs, an example of the latter being a PSR orbiting Sgr~A*. We will use simulations of the fractional measurement accuracy $\delta_\Q$ of the BH quadrupole moment for these two cases respectively. First, Ref.~\cite{WexBHP} simulates a stellar-mass BH quadrupole moment measurability in a binary with a millisecond PSR as a function of BH mass. The reference uses the following system parameters: eccentricity $e = 0.5$, 10 Myr merger lifetime, either $20^\circ$ or $45^\circ$ spin inclination angle, orbital period of either 0.16 or 0.21 days, $1.4 \Msolar$ PSR mass, a $60^\circ$ angle between line of sight and the same observation scheme.
Second, Ref.~\cite{QuadrupoleMeasureabilitySgrA_PsaltisWex} simulates the measurability of the Sgr A* quadrupole moment by timing a PSR orbiting it. The reference assumes a 100 $\mu$s TOA precision measured three times daily for 3 periastron or full orbit passages achieved by observation with SKA or the 100m telescopes. The simulation uses a BH spin of 0.36, eccentricity $e=0.8$ and an orbital period of 0.5 yr. Finally, the simulated fractional measurement accuracy of the BH quadrupole moment is presented in Fig.~\ref{fig:Quad_Mom_Measure}. Below we will use this measurement accuracy of $Q$ to derive projected bounds on quadratic-curvature corrected theories.

\subsection{Example Theories and Projected Bounds} 
\label{sec:Q-example}

We now study how well one can probe specific modified theories of gravity with the BH quadrupole moment measurement via BH-PSRs with SKA. We will study two example theories, both of which modify the Einstein-Hilbert action by introducing a correction term, in which a (pseudo-)scalar field is coupled to curvature-squared in a non-minimal way. Such theories are motivated as low-energy effective theories of string theory.
EdGB gravity represents the parity-even sector of such quadratic gravity, while dCS gravity represents its odd-parity sector. The latter is of particular interest here since corrections to $\dot P$ enters first at 2PN order. Thus, $Q$ measurements have more advantage on probing dCS than $\dot P$ measurements. We will also study EdGB gravity to demonstrate that theories with negative PN corrections to $\dot P$ do not benefit from $Q$ measurements over those for $\dot P$.

\subsubsection{Dynamical Chern-Simons Gravity}\label{subsec:DynamicalCS}
DCS is a parity violating theory of gravity in which a pseudoscalar field is coupled to the Pontryagin density given by ${}^*R_{\mu\nu\rho\sigma}R^{\mu\nu\rho\sigma}$, where $R^{\mu\nu\rho\sigma}$ is the Riemann tensor while ${}^*R^{\mu\nu\rho\sigma}$ is its dual~\cite{Jackiw:2003pm,Alexander:2009tp}. This theory not only arises from string theory but also from loop quantum gravity~\cite{Ashtekar:1988sw,alexandergates,Taveras:2008yf,calcagni,Gates:2009pt}, chiral anomaly in the Standard Model~\cite{Weinberg:1996kr} and effective theories of inflation~\cite{Weinberg:2008hq}. Since the field equations contain third derivatives, one needs to treat the theory as an effective field theory within the small coupling approximation to keep the theory well-posed~\cite{Delsate:2014hba}\footnote{See~\cite{Cayuso:2017iqc} for another way to possible cure higher derivative pathologies in the theory.}. 

BHs in dCS gravity acquire deviations from GR only when they are spinning. Slowly-rotating solutions have been constructed to first order~\cite{Yunes:2009hc,Konno:2009kg}, second order~\cite{DCSBoundSpin2} and fifth order~\cite{DCSBoundspin4} in spin. Pseudoscalar field configuration has been studied semi-analytically~\cite{Konno:2014qua} and numerically~\cite{Stein:2014xba} for arbitrary spin, and for extremal BHs~\cite{McNees:2015srl}. Delsate \textit{et al.}~\cite{Delsate:2018ome} recently constructed rapidly-rotating BH solutions numerically. In this paper, we adopt the BH quadrupole moment within the small coupling and slow rotation approximation, valid to first order in dimensionless coupling constant and to fourth order in spin~\cite{DCSBoundSpin2,DCSBoundspin4}\footnote{This expression is actually valid to fifth order in spin since the next spin correction enters at $\mathcal{O}(\chi^6)$~\cite{DCSBoundspin4}.}:

\begin{equation}
Q = Q_\GRk\left(1-\frac{201}{1792} \zeta_\dCS +\frac{1819}{56448} \zeta_\dCS \chi ^2\right)\;.
\label{eq:DCSQuadrupole}
\end{equation}
Here
\begin{equation}
\label{eq:zeta-dCS}
\zeta_\dCS = \frac{\alpha_\dCS^2}{\kappa_g m_c^4} 
\end{equation}
is the dimensionless coupling constant with $\kappa_g = (16\pi)^{-1}$, $m_\BH$ being the BH mass and $\alpha_\dCS$ representing the dimensional coupling constant (in units of length squared). The dimensionless spin is defined by $\chi =J/m_\BH^2$, with $J$ representing the magnitude of the spin angular momentum and $Q_\GRk = - M^3 \chi^2$ is the quadrupole moment of a Kerr BH~\cite{hansen}. Using the following relation in the uncertainty of the quadrupole moment,
\begin{equation}
\label{eq:deltaQ}
\left| \frac{Q- Q_\GRk}{Q_\GRk}\right| \leq \delta_\Q \;,
\end{equation}
we can solve for $\alpha_\dCS^{1/2}$ as 
\begin{equation}
\alpha_\dCS^{1/2}\leq4 \sqrt{21} \left(\frac{\kappa _g \, \delta _\Q}{12663
 -3638 \chi ^2} \right)^{1/4} m_\BH\;. \label{eq:dCSalphabound}
\end{equation}

\begin{figure}[!h]
\begin{center}
\includegraphics[width=\linewidth]{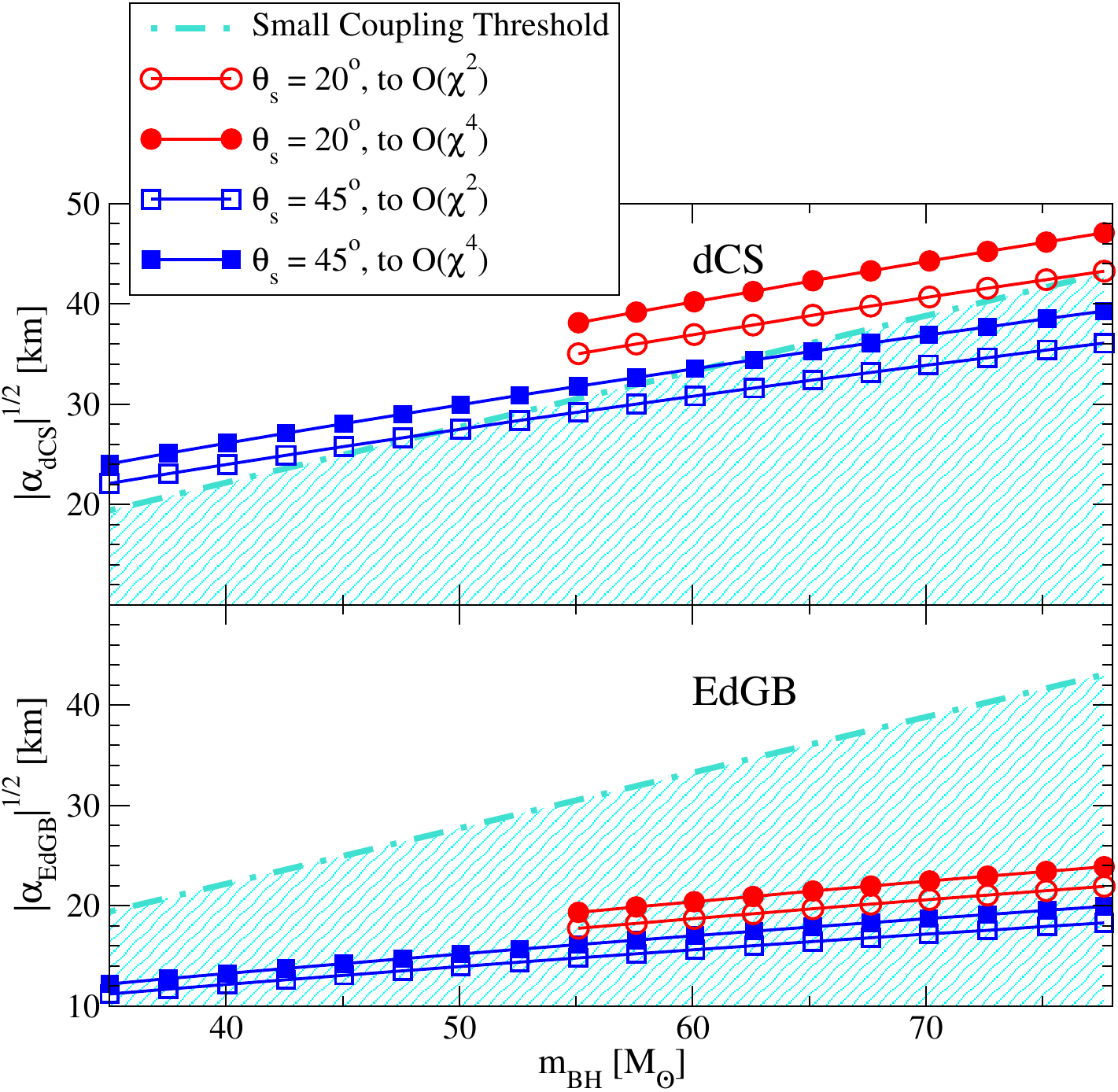}
\caption{\label{fig:dCSEdGBBoundofMass} 
Projected upper bounds on the square root of the coupling constant in dCS (top) and EdGB (bottom) gravity assuming that SKA measures the BH quadrupole moment in stellar-mass BH-PSR binaries, as a function of the companion BH mass. We derive the bounds by truncating the BH quadrupole moment in Eqs.~\eqref{eq:DCSQuadrupole} and~\eqref{eq:Q-EdGB} to $\mathcal{O}(\chi^2)$ (open circles or squares) and by using the full expression valid to $\mathcal{O}(\chi^4)$ (filled circles or squares). We consider two different spin inclination angles $\theta_s$ of $ 20^\circ$ (red circle) and $ 40^\circ$ (blue square). Other binary parameters are the same as those for the left panel in Fig.~\ref{fig:Quad_Mom_Measure}. In particular, since the fiducial spin was assumed to be $\chi = 1$, the slow-rotation approximation adopted in Eqs.~\eqref{eq:DCSQuadrupole} and~\eqref{eq:Q-EdGB} is not accurate, though the bounds are valid as an order of magnitude estimate. The allowed regions for the small coupling approximation are shaded. The BH-PSR bounds are much stronger than the current strongest bounds from solar system~\cite{AliHaimoud:2011fw} and table-top~\cite{DCSBoundSpin2} experiments ( $\, \sqrt[]{\alpha_\dCS} = \mathcal{O}(10^8)$ km) for dCS gravity, while they are much weaker than bounds from BH-LMXBs ($\, \sqrt[]{\alpha_\EdGB} = 1.9$ km)~\cite{Yagi:2012gp} for EdGB gravity. 
}
\end{center}
\end{figure}
Let us first study projected upper bounds on dCS gravity with stellar-mass BH-PSRs.
The top panel of Fig.~\ref{fig:dCSEdGBBoundofMass} presents such bounds on $\sqrt{\alpha_\dCS}$ from the BH quadrupole moment measurement with stellar-mass BH-PSRs using SKA, as a function of the BH mass. This figure is obtained by using Eq.~\eqref{eq:deltaQ} with $\delta_\Q$ shown in the left panel of Fig.~\ref{fig:Quad_Mom_Measure}. We use two different spin inclination angles and compare the cases where we truncate Eq.~\eqref{eq:DCSQuadrupole} at $\mathcal{O}(\chi^2)$ and where we use its full expression valid to $\mathcal{O}(\chi^4)$. Observe that the latter two cases differ by $\sim 10\%$. This is because the fiducial value for the BH spin in the left panel of Fig.~\ref{fig:Quad_Mom_Measure} is $\chi = 1$, and the slow-rotation approximation imposed in Eq.~\eqref{eq:DCSQuadrupole} is not strictly valid. Since the difference between open and filled circles/squares in Fig.~\ref{fig:dCSEdGBBoundofMass} is relatively small, we conclude that the projected bounds are valid as an order of magnitude estimate. 

Are projected bounds in Fig.~\ref{fig:dCSEdGBBoundofMass} weak or strong?
Comparing such projected bounds from BH-PSRs to current strongest bounds of $\mathcal{O}(10^8)$km obtained from solar system~\cite{AliHaimoud:2011fw} and table-top~\cite{DCSBoundSpin2} experiments, we see that the former are stronger than the latter by six to seven orders of magnitude in Fig.~\ref{fig:dCSEdGBBoundofMass}. Note that because Eq.~\eqref{eq:dCSalphabound} depends on $\delta_\Q$ to the power of $1/4$, the analysis here is insensitive to improvements in quadrupole moment measurability. Thus, it is a promising method for bounding dCS due to the relative independence of measurement precision. We also note that NS-PSR/PSR-WD binaries are not efficient in constraining dCS gravity because NS spins are too small to obtain any meaningful bounds on the theory~\cite{DCSKent}. 

The correction to the BH quadrupole moment in Eq.~\eqref{eq:DCSQuadrupole} has been derived within the small coupling approximation $\zeta_\dCS \ll 1$, and thus we need to check whether the bounds from BH-PSRs in Fig.~\ref{fig:dCSEdGBBoundofMass} are valid within such an approximation. The above condition for the small coupling approximation can be rewritten as
\begin{equation}\label{eq:smallcoupling}
\alpha_\dCS^{1/2} \ll \kappa_g^{1/4} m_\BH \; .
\end{equation}
The shaded region in Fig.~\ref{fig:dCSEdGBBoundofMass} represents the parameter space in which the small coupling approximation is valid. Observe that the BH-PSR bounds are only meaningful for $\theta_s \sim 45^\circ$ and $m_\BH \gtrsim 50M_\odot$.

Next, let us examine bounding dCS gravity from supermassive BH-PSRs. The small coupling threshold for $\alpha\dCS^{1/2}$ is equal to $3\times10^{6}$ km. The strongest possible bound here comes from full orbit measurements from a PSR orbiting Sgr A*. Unfortunately, such strongest bound is above the small coupling threshold of $\alpha_\dCS^{1/2}$ by about 20\%. The measurements of periapsis only are even weaker, so they do not satisfy the small coupling requirement. Due to the the fact that $\alpha_\dCS^{1/2}\propto\delta_\Q^{1/4}$, better measurement accuracy is unlikely to allow the bound to satisfy the small coupling threshold. This analysis suggests that quadrupole moment modification in a PSR-Sgr A* binary will be unlikely to produce an improved bound for dCS.

\subsubsection{Einstein-dilaton Gauss-Bonnet Gravity}\label{subsec:EdGB}
Similar to dCS, EdGB gravity also introduces a curvature-squared term in the action. Such a term is the Gauss-Bonnet invariant which is non-minimally coupled to a scalar field, and thus the theory is parity even. String theory predicts that the coupling between scalar field and gravity is given in an exponential form but in this paper, we consider a linear coupling. This theory is referred to as decoupled dynamical Gauss-Bonnet gravity in~\cite{EDGBBoundComparison} and corresponds to Taylor expanding the scalar field about some constant and keeping up to linear order in the scalar field\footnote{The term where the scalar field does not couple to the Gauss-Bonnet invariant is a total derivative and does not contribute to the field equations}. 

BH solutions in linearly-coupled Gauss-Bonnet gravity have been constructed analytically for both static~\cite{Yunes:2011we,Sotiriou:2014pfa} and slowly-rotating configurations~\cite{EDGBBoundSpin2,EDGBBoundSpin4}. In this theory, BHs acquire monopole scalar charges~\cite{Yunes:2011we,Sotiriou:2014pfa} that produce scalar dipole radiation in BH binaries~\cite{Yagi:2011xp}. Such radiation modifies $\dot P$ at $-1$PN order. On the other hand, ordinary stars like neutron stars do not possess such scalar charges~\cite{Yagi:2011xp,EDGBBoundComparison}, and thus binary PSRs cannot place stringent bounds on the theory. One of the most stringent bounds in this theory has been obtained from the orbital decay rate measurement of a BH low-mass X-ray binaries as $\sqrt{\alpha_\EdGB} < 1.9 $km~\cite{Yagi:2012gp}, where $\alpha_\EdGB$ is the dimensional coupling constant in the theory. Similar bounds have been derived in EdGB gravity from the existence of stellar-mass BHs~\cite{Kanti:1995vq,Pani:2009wy} and the maximum mass of neutron stars~\cite{Pani:2011xm}, though the latter depends on the unknown equation of state for neutron stars. Stronger bounds can be obtained from $\dot P$ measurements with BH-PSRs using FAST or SKA~\cite{EDGBBoundComparison}. 

We now review the BH quadrupole moment in EdGB gravity that modifies the Roemer time delay from GR. The quadrupole moment valid to fourth order in spin within the small coupling approximation is given by~\cite{EDGBBoundSpin2,EDGBBoundSpin4}
\begin{equation}
\label{eq:Q-EdGB}
 Q= Q_\GRk\left(1+\frac{4463}{2625}\zeta_\EdGB-\frac{33863}{68600} \zeta_\EdGB \chi ^2 \right) \;,
\end{equation}
where $\zeta_\EdGB$ is the dimensionless coupling constant defined by Eq.~\ref{eq:zeta-dCS} but replacing $\alpha_\dCS$ with $\alpha_\EdGB$. Substituting the above equation to Eq.~\eqref{eq:dCSalphabound} and solving for $\alpha_\EdGB^{1/2}$, one finds

\begin{equation}
\alpha_\EdGB^{1/2}\leq 3^{1/4} \, 70^{3/4} \left( {\frac{\kappa _g \delta
 _\Q}{1749496-507945 \chi ^2}} \right)^{1/4}m_\BH\; .
\end{equation}

The bottom panel of Fig.~\ref{fig:dCSEdGBBoundofMass} presents projected upper bounds on $\sqrt{\alpha_\EdGB}$ for stellar-mass BH-PSRs from the BH quadrupole moment measurement with SKA as a function of the BH mass. Observe that such bounds are much weaker than the bounds from a BH-LMXB mentioned earlier. This is because the orbital decay rate measurement has more advantage on probing theories that give rise to $-1$PN corrections to $\dot P$ than the quadrupole moment measurement. We do not present projected bounds from the quadrupole moment measurement of Sgr~A* with PSRs since such bounds would be even weaker than those in Fig.~\ref{fig:dCSEdGBBoundofMass}. Unlike the dCS case, BH-PSR bounds on EdGB gravity can easily satisfy the small coupling approximation. This is because the quadrupole moment correction is larger for EdGB gravity than in dCS gravity (compare the coefficients in front of $\zeta$ in Eqs.~\eqref{eq:DCSQuadrupole} and~\eqref{eq:Q-EdGB}).

\section{Conclusion and Discussion}
\label{sec:conclusion}

In this paper, we studied how well one can probe gravity with orbital decay rate and BH quadrupole moment measurements using future BH-PSR observations. Regarding the former, we showed that bounds from generic non-GR modifications to $\dot P$ can be stronger than those from the double PSR by a few orders of magnitude, especially for corrections entering at negative PN orders. We mapped this result to various example modified theories of gravity and found that e.g.~bounds on $\dot G$ can be much stronger than the current bound from solar system experiments. Regarding the latter, we showed that the Roemer time delay for certain stellar-mass BH-PSR configurations can be used to place bounds on dCS gravity that are six orders of magnitude stronger than the current most stringent bounds. Thus, the detection of a BH-PSR binary will allow new tests of gravity.


Future work includes extending BH-PSR bounds to Lorentz-violating theories, such as Einstein-\ae ther~\cite{Jacobson:2000xp,Jacobson:2008aj} and khronometric~\cite{Blas:2009qj,Blas:2010hb} gravity. Scalar and vector degrees of freedom in those theories induce dipole radiation that depends on the sensitivities of compact objects. So far, such sensitivities for strongly-gravitating objects have been calculated for NSs~\cite{Yagi:2013qpa,Yagi:2013ava}. One can repeat the analysis in~\cite{Yagi:2013qpa,Yagi:2013ava} by constructing slowly-moving BH solutions with respect to the vector field in these theories to extract BH sensitivities within the parameter space allowed from theoretical and observational constraints including GW170817~\cite{Gumrukcuoglu:2017ijh,Oost:2018tcv}. One can combine this with the orbital decay rate in these theories derived in~\cite{Yagi:2013ava} to estimate projected bounds on such theories from future BH-PSR observations.

Another avenue for future work includes improving the analysis for bounding dCS gravity from BH-PSR observations. In this paper, we used the stellar mass BH quadrupole moment obtained within the slow-rotation approximation~\citep{DCSBoundSpin2,DCSBoundspin4} for BH-PSR systems with the BH dimensionless spin of unity, and thus the bounds should only be understood as order of magnitude estimates. One could improve this analysis by deriving the BH quadrupole moment valid for arbitrary spin. Such a goal may be achieved by using BH solutions with arbitrary spin in dCS gravity recently constructed numerically~\cite{Delsate:2018ome}. 

It would also be interesting to derive bounds on dCS gravity from the measurement of the advance rate of periastron for BH-PSRs. This is because Ref.~\cite{DCSKent} identified such an observable to be a useful post-Keplerian parameter when bounding dCS gravity from binary PSR observations. Typically masses are measured with e.g.~the advance rate of periastron and Shapiro time delay, and thus if GR tests are done with advance rate of periastron, one must use a different PPK parameter to derive the masses. This approach introduces more uncertainty to the bound because the mass measurement precision is reduced. Ideally, one needs to search for non-GR parameters and determine the masses simultaneously, instead of using the derived masses assuming GR is correct.

\acknowledgments
We would like to thank Norbert Wex for giving us helpful comments.
K.Y. acknowledges support from NSF Award PHY-1806776. 
K.Y. would like to also acknowledge networking support by the COST Action GWverse CA16104.

\appendix 
\section{Correction for an Integral Formula in Cubic Galileon }\label{ap:DipoleCGCorrection}
In \cite{MassiveGravityCG}, we noticed that Eq.~(4.10) had a typo, which we correct in this appendix. The integral that is relevant for evaluating dipolar radiation in cubic Galileon massive gravity is given by
\begin{equation}
I_k^D(e) = (1-e^2)^3\frac{k^{13/4}}{2\pi}\int_0^{2\pi}\frac{\exp[-i(k-1)x]}{(1+e \cos{x})^3}dx \;,
\end{equation}
where $k$ is an integer with $k \geq 0$.
First, notice that symmetry over $0$ to $2\pi$ allows the following simplification,
\begin{equation}\label{eq:DipInt}
\int_0^{2\pi}\frac{\exp[-i(k-1)x]}{(1+e \cos{x})^3}dx = 2\int_0^{\pi}\frac{\cos{[(k-1)x]}}{(1+e \cos{x})^3}dx\;.
\end{equation}

On the other hand, one finds the integral formula from Eq.~(3.613) of \cite{Gradshtein2007TableProducts} for $e^2 < 1$ and $m\geq 0$:
\begin{equation}\label{eq:DipTabEntry}
\int_0^{\pi}\frac{\cos{(m x)}}{1+e \cos{x}}dx=\frac{\pi}{\sqrt[]{1-e^2}}\left(\frac{\sqrt[]{1-e^2}-1}{e}\right)^m \;.
\end{equation}
Using Feynman's trick to reduce the power in the denominator of the integrand in the right-hand-side of Eq.~\eqref{eq:DipInt} to unity, we can use the above integral formula as 
\begin{widetext}
\begin{eqnarray}
2\int_0^{\pi}\frac{\cos{[(k-1)x]}}{(1+e \cos{x})^3}dx &=&2 \lim_{b \to 1}\int_0^{\pi}\frac{1}{2}\partial_b^2\frac{\cos{[(k-1)x]}}{b+e \cos{x}}\nonumber \\
&= &\lim_{b \to 1}\partial_b^2 \frac{1}{b}\int_0^{\pi}\frac{\cos{[(k-1)x]}}{1+\frac{e}{b} \cos{x}} \ = \ \lim_{b \to 1}\partial_b^2 \frac{\pi}{\sqrt[]{b^2-e^2}}\left(\frac{\sqrt[]{b^2-e^2}-b}{e}\right)^{k-1}\; .
\end{eqnarray}

Thus, $I_0^D(e) = 0$, and $I_k^D(e)$ with $k \geq 1$ becomes 
\begin{eqnarray}\label{eq:DipIntEval}
I_k^D(e) &=&(1-e^2)^3\frac{k^{13/4}}{2\pi} \lim_{b \to 1}\partial_b^2 \frac{\pi}{\sqrt[]{b^2-e^2}}\left(\frac{\sqrt[]{b^2-e^2}-b}{e}\right)^{k-1} \nonumber \\
&=&\frac{1}{2} k^{13/4} \sqrt{1-e^2}\left(\frac{\sqrt{1-e^2}-1}{e}\right)^{k-1}\left[ 3-3 \sqrt{1-e^2}+\left(2 e^2+3 \sqrt{1-e^2}-2\right) k +\left(1-e^2\right) k^2\right] \; .\label{eq:IDipole}
\end{eqnarray}
\end{widetext}
Comparing this with Eq.~(4.10) in \cite{MassiveGravityCG}, we realized that the typo is that the latter has $(\sqrt[]{1-e^2}-1)^{k-1}$ instead of $[(\sqrt[]{1-e^2}-1)/e]^{k-1}$.
With Eq.~\eqref{eq:IDipole}, the sum of $|I_k^D(e)|^2$ in the dipole radiation formula (Eq.~(4.8) of \cite{MassiveGravityCG}) can be more accurately calculated without numerical integration.

\section{Comparison of Bounds from GW170817 with those from Binary Pulsars }\label{ap:GW170817}

For the bound on $\gamma$ in the main text, we used analysis from Ref.~\cite{Yunes:2016jcc} for the two gravitational wave detections GW150914 and GW151226 consistent with binary BH mergers \cite{Abbott:2016blz,Abbott:2016nmj}. Since that paper has been published, there was the detection of GW170817 consistent with a binary NS merger \cite{TheLIGOScientific:2017qsa}, which is able to place stronger bounds than the previous two. In this appendix, we will be showing why binary PSRs still have an improvement compared to GW bounds at low PN orders.

\begin{figure}[!h]
\begin{center}
\includegraphics[width=\linewidth]{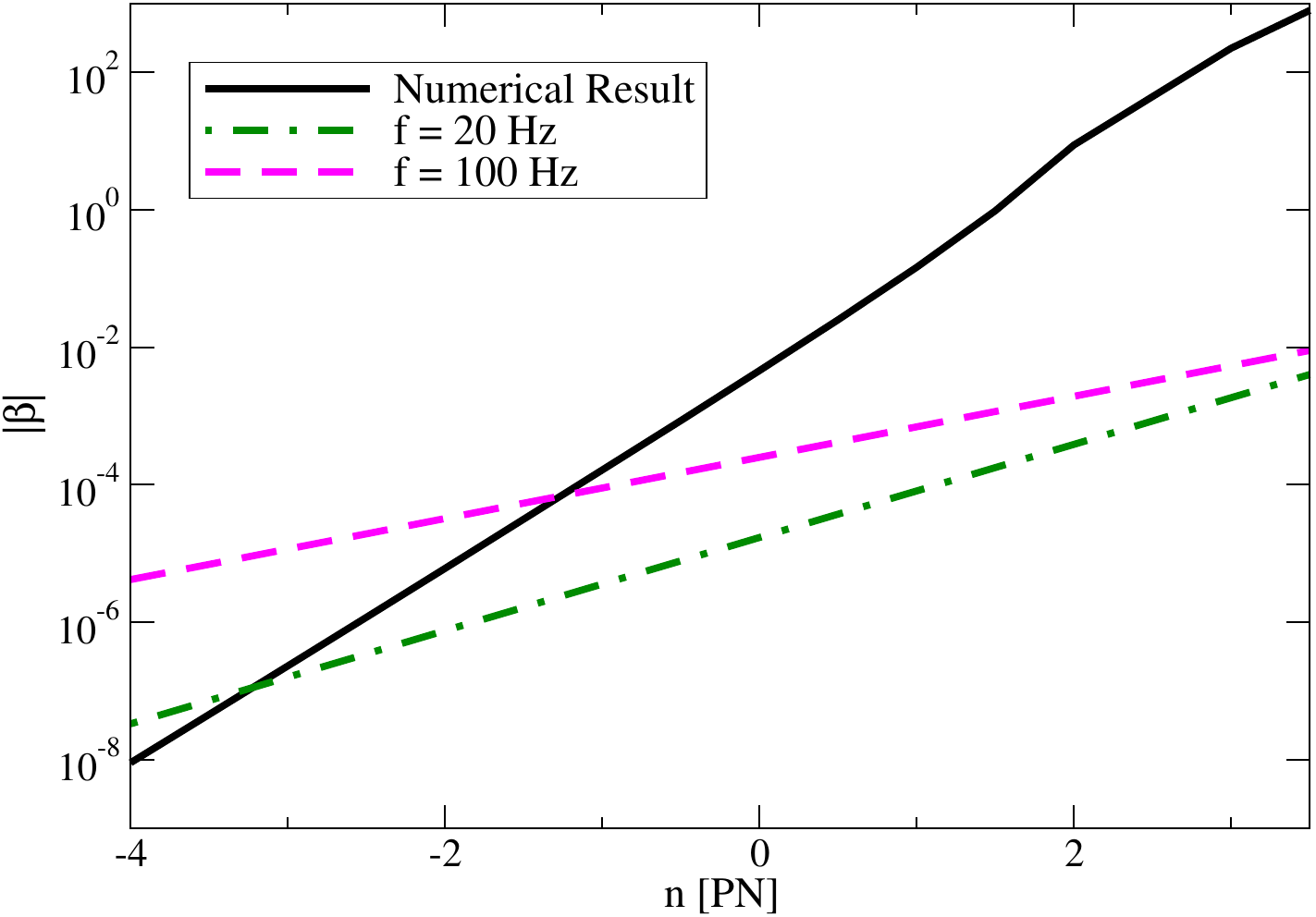}
\caption{\label{fig:betacomparison} 
Crude estimate on the upper bound on the PPE parameter $\beta$ for GW150914 as a function of at which PN order the correction enters, for two representative GW frequencies (20 Hz and 100 Hz). For reference we also present the bounds using a numerical Fisher analysis~\cite{Yunes:2016jcc}. Observe that the two crude estimates bracket the numerical one at lower PN order.}
\end{center}
\end{figure}

What must be determined is the upper bound on $\beta$ for a gravitational wave detection. First, we will motivate our crude estimate by comparing it with the known bound from GW150914 using a Fisher analysis~\cite{Yunes:2016jcc}. We estimate the bound on $\beta$ in the following way \cite{Cornish:2011ys,Hayasaki:2012qn}: 
\begin{equation}
\delta \Phi(f)=\beta (\pi \mathcal{M} f)^{b/3} \lesssim \frac{1}{\rho} \; , \label{eq:BetaGW} \;
\end{equation}
where $\mathcal{M}$ is the chirp mass while $\rho$ is the signal-to-noise ratio (SNR). 
For GW150914, $\mathcal{M} = 28.1M_\odot$ and $\rho = 23.7$ \cite{TheLIGOScientific:2016pea}.
The reason why the above equation roughly holds is because a GW measurement can distinguish two waveforms with a phase offset of $\mathcal{O}(1~\mathrm{rad})/\mathrm{SNR}$. In Fig.~\ref{fig:betacomparison}, we show this crude estimate compared to the Fisher result in~\cite{Yunes:2016jcc} for two representative choices of the GW frequency. One sees that $\beta$ calculated at 20 Hz and 100 Hz brackets the range of the true value for PN order less than $-1$. 20 Hz corresponds to the minimum frequency that Advanced LIGO is sensitive to~\cite{TheLIGOScientific:2016src}, while 100Hz is roughly the frequency that the detector is most sensitive. On the other hand, our crude estimate does not work for higher PN corrections. This is because for such cases, non-GR parameters have a huge degeneracy with other binary parameters like masses and spins, which is not taken into account in our estimate.. This result motivates a possible crude estimation scheme for GW170817 for lower PN order bounds. For the remaining of this appendix, we focus on the PN order lower than $-1$.

\begin{figure}[!h]
\begin{center}
\includegraphics[width=\linewidth]{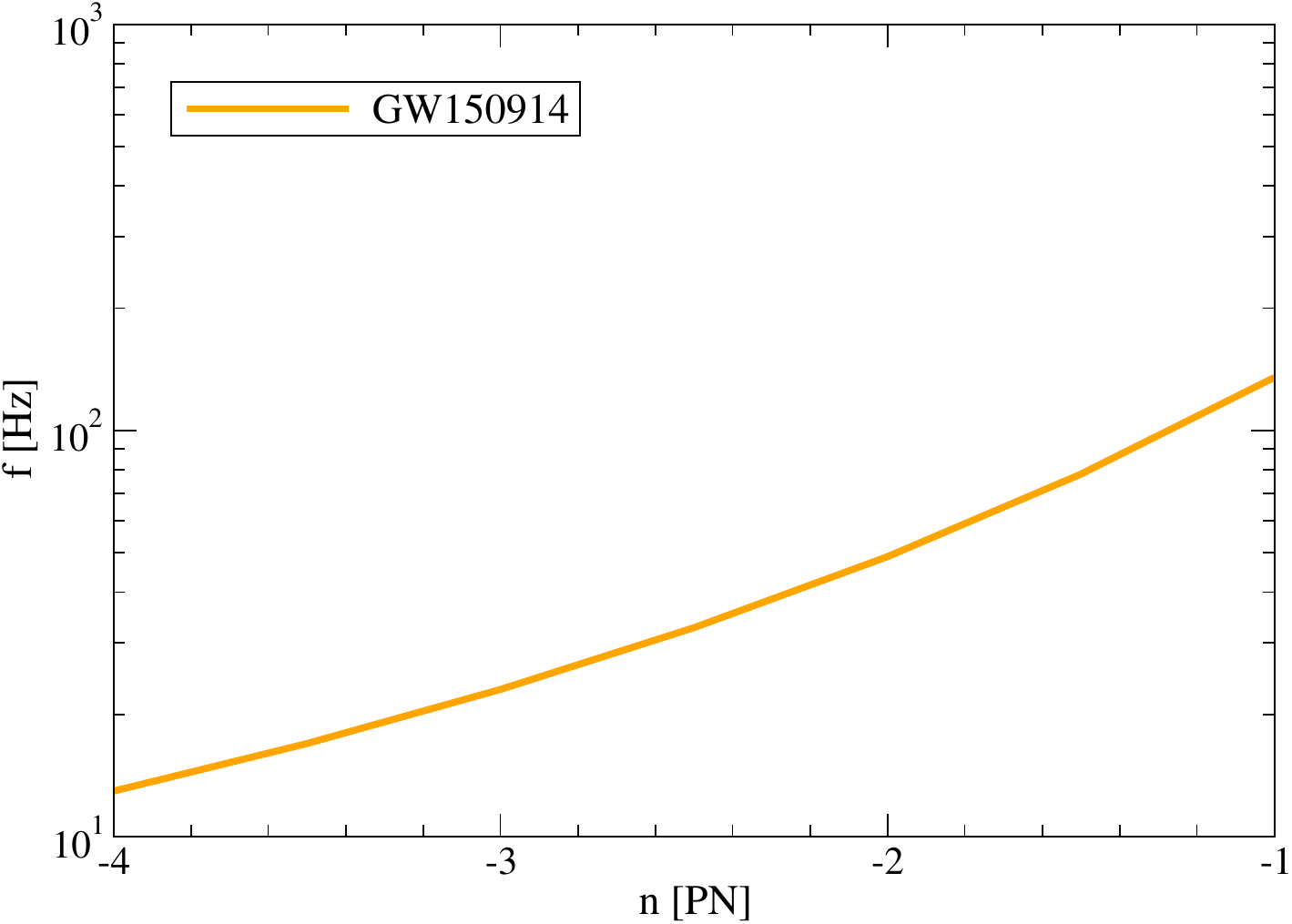}
\caption{\label{fig:frequencies} 
The GW frequency of GW150914 as a function of PN such that the cruder estimate in Eq.~\eqref{eq:BetaGW} becomes exactly the same as that obtained from the Fisher analysis in~\cite{Yunes:2016jcc}. Observe that such dominant GW frequency becomes lower for lower PN orders corrections.
}
\end{center}
\end{figure}

Figure~\ref{fig:betacomparison} clearly shows that the GW frequency that gives the dominant contribution in terms of bounding $\beta$ changes for each PN order, and a lower frequency contribution becomes larger for lower PN order corrections. In order to find such dominant GW frequency at each PN order, we solve Eq.~\eqref{eq:BetaGW} for frequency with $\beta$ fixed to the upper bound obtained from a Fisher analysis, which we show in Fig.~\ref{fig:frequencies}. In other words, if one uses the frequency shown in Fig.~\ref{fig:frequencies} at each PN order in the crude estimate expression of Eq.~\ref{fig:frequencies}, one can recover the upper bound on $\beta$ obtained from the Fisher calculation.

\begin{figure}[!h]
\begin{center}
\includegraphics[width=\linewidth]{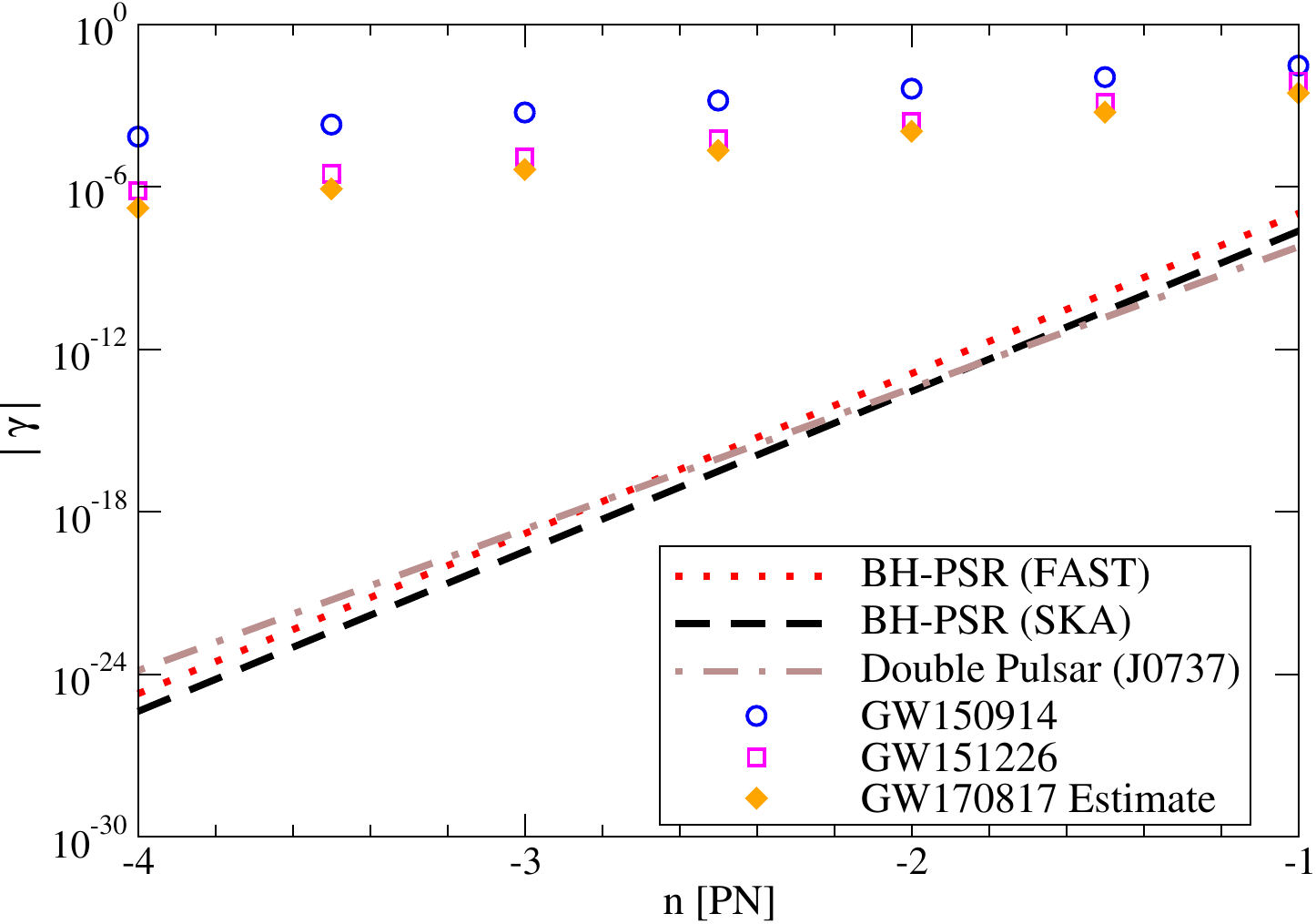}
\caption{\label{fig:gammapnGW170817} 
The upper bound on the fractional non-GR correction to the orbital decay rate $\gamma$ in Eq.~\eqref{eq:gammaboundvb} at each PN order for various astrophysical systems. This is a replication of Fig.~\ref{fig:gammapn} at low PN order with the addition of the speculative bound of GW170817. Note that the GW170817 bound on $\gamma$ is not close to being competitive with those from BH/PSRs and the double pulsar.
}
\end{center}
\end{figure}

We are now ready to estimate the bounds on $\beta$ (or equivalently $\gamma$) from GW170817. We use Eq.~\eqref{eq:BetaGW} but use $\mathcal{M} = 1.188M_\odot$, $\rho = 32.4$ \cite{TheLIGOScientific:2017qsa} and $f$ in Fig.~\ref{fig:frequencies}. We then map the bound on $\beta$ to that on $\gamma$ using Eq.~\eqref{eq:gamma-beta}. We present the result in Fig.~\ref{fig:gammapnGW170817}, together with other bounds from Fig.~\ref{fig:gammapn}. Observe that the GW170817 bounds are stronger than those from GW150914 and GW151226, though such bounds are not even close to those with BH/PSRs and the double pulsar.

\bibliography{Bibliography.bib}

\end{document}